\documentclass[conference]{IEEEtran}
\IEEEoverridecommandlockouts
\usepackage{cite}
\usepackage{amsmath,amssymb,amsfonts}
\usepackage{algorithmic}
\usepackage[dvips]{graphicx}
\usepackage{textcomp}
\usepackage{xcolor}
\PassOptionsToPackage{hyphens}{url}\usepackage{hyperref}
\def\BibTeX{{\rm B\kern-.05em{\sc i\kern-.025em b}\kern-.08em
    T\kern-.1667em\lower.7ex\hbox{E}\kern-.125emX}}
\begin{document}

\title{The Impact of COVID-19 on Flight Networks}

\author{
\IEEEauthorblockN{*Toyotaro Suzumura\IEEEauthorrefmark{1}, *Hiroki Kanezashi\IEEEauthorrefmark{2}, Mishal Dholakia\IEEEauthorrefmark{3}, Euma Ishii\IEEEauthorrefmark{4,5},\\ Sergio Alvarez Napagao\IEEEauthorrefmark{1}, Raquel Pérez-Arnal\IEEEauthorrefmark{1} Dario Garcia-Gasulla\IEEEauthorrefmark{1} and Toshiaki Murofushi\IEEEauthorrefmark{2}}
    \IEEEauthorblockA{\IEEEauthorrefmark{1}Barcelona Supercomputing Center
    \\ suzumura@acm.org, sergio.alvarez@bsc.es, raquel.perez@bsc.es, dario.garcia@bsc.es}
    \IEEEauthorblockA{\IEEEauthorrefmark{2}Tokyo Institute of Technology
    \\ kanezashi.h.aa@m.titech.ac.jp, murofusi@c.titech.ac.jp}
    \IEEEauthorblockA{\IEEEauthorrefmark{3}IBM
    \\ mishal.dholakia1@ibm.com}
    \IEEEauthorblockA{\IEEEauthorrefmark{4}Massachusetts Institute of Technology
    \\ euma@mit.edu}
    \IEEEauthorblockA{\IEEEauthorrefmark{5}Tokyo Medical and Dental University
    \\ 152011ms@tmd.ac.jp}
    \thanks{*Toyotaro Suzumura and Hiroki Kanezashi contributed equally to this work.}
}

\maketitle

\begin{abstract}
As COVID-19 transmissions spread worldwide, governments have announced and enforced travel restrictions to prevent further infections. Such restrictions have a direct effect on the volume of international flights among these countries, resulting in extensive social and economic costs. To better understand the situation in a quantitative manner, we used the OpenSky Network data to clarify flight patterns and flight densities around the world and observe relationships between flight numbers with new infections, and with the economy (unemployment rate) in Barcelona.
We found that the number of daily flights gradually decreased and suddenly dropped 64\% during the second half of March in 2020 after the US and Europe enacted travel restrictions. We also observed a 51\% decrease in the global flight network density decreased during this period. Regarding new COVID-19 cases, the world had an unexpected surge regardless of travel restrictions. Finally, the layoffs for temporary workers in the tourism and airplane business increased by 4.3 fold in the weeks following Spain's decision to close its borders.
\end{abstract}

\section{Introduction}

\subsection{Background}
As COVID-19 transmissions spread worldwide, governments announced and enforced both domestic and international travel restrictions. These restrictions affected the trend of flight networks around the world such as major airlines \cite{whitehouse-airlines} and tourism-dependent cities \cite{tourism-gdp}. All or a portion of flights connecting these restricted countries and areas were cancelled. The date of travel restriction enforcement varies by country, therefore the period and degree of flight reductions also vary by country and continents.

\subsection{Overview}
We quantitatively investigated how flight restrictions affect domestic and international travel in countries and continents through time-series analytics and graph algorithms. We also investigated the negative effect of these restrictions on employment and the number of infected cases in Barcelona, where the majority of revenue depends on the travel industry.

The structure of this paper is as follows; In Section \ref{sec:methods}, we introduce the flight dataset which we used for our analysis. We compared the dataset with other flight datasets and discussed the limitations and potential biases. In Section \ref{sec:stat}, we summarize the statistical analytics of airport and flight data by country and continent. In Section \ref{sec:timeseries}, we analyze the number of daily flights globally as well as by region (continent and country level), and discuss the relationship between government announcements and the decline of flights. In Section \ref{sec:graph}, we visualize flight networks and apply graph analytic techniques to the network to evaluate the quantitative impacts on travelers. We examine economic effects on travel restrictions and cities depending on tourism industries from the public data from Barcelona in Section \ref{sec:barcelona}, and we evaluate the relationship between incoming  flights from Europe and the number of new infection cases in the United States in Section \ref{sec:cases}. Finally, we discuss our findings, limitations and related work in Section \ref{sec:discussion}, and explain the summary and future perspectives of our research in Section \ref{sec:conclusion}.
\section{Methods}
\label{sec:methods}

\subsection{Methods and Algorithms}
In order to evaluate the effect of travel restrictions on the flights, we conducted the following analyses step by step.

First, we summarized the total number of airports, flights and potential passengers for each country and continent from an open dataset.
Next, we counted up the number of daily flights for countries and continents, and compared the period of the flight reductions and the date of the travel restriction enforcement.
In order to evaluate the density of the flight network quantitatively, we also applied graph analytics methods to the time-series flight data.
Finally, we analyzed time-series data sets about the number of incoming flights, number of infected cases and unemployment in Barcelona \cite{barcelona-tourism}, where tourism industries is a main revenue source, and hypothesized the correlation among these data.

\subsection{The OpenSky Network Dataset}
For time-series flight analytics, we used an open dataset \cite{opensky-data} from The OpenSky Network\cite{opensky-web}. The dataset contains flight records with departure and arrival times, airport codes (origin and destination), and aircraft types.
The dataset includes the following flight information during 2020 January 1 to April 30. The dataset for a particular month is made available during the beginning of the following month. Note that the year of all of the following dates is 2020. In total, our dataset is composed by:
\begin{itemize}
  \item 655,605 (35\%) international flights
  \item1,213,100 (65\%) domestic flights
  \item 616 major airports covering 148 countries (out of 195 world countries)
\end{itemize}

\section{Statistical Analysis of Flight Data}
\label{sec:stat}

\subsection{Distribution of the Number of Airports}
\begin{figure}[htb]
\centerline{\includegraphics[width=\hsize]{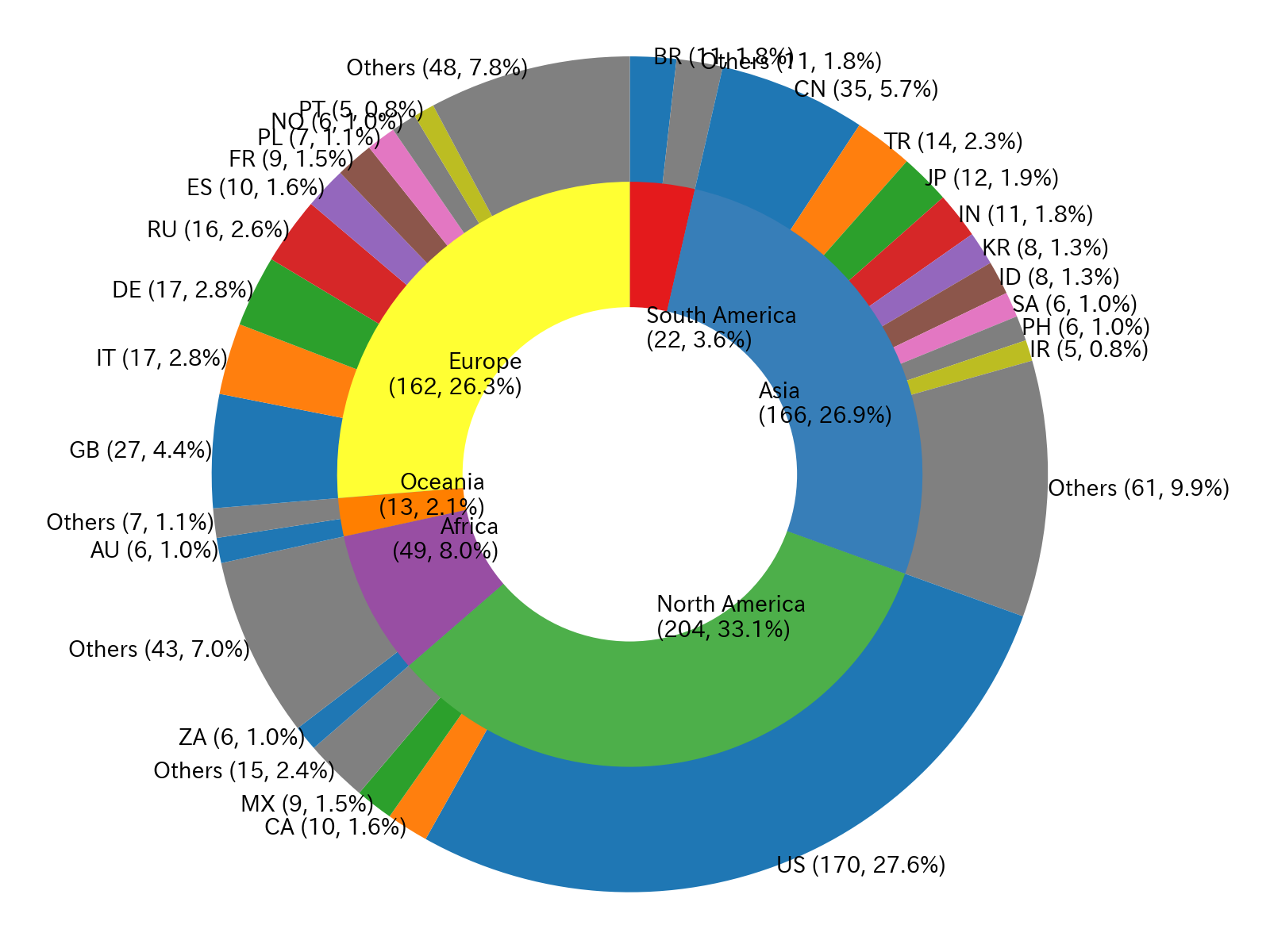}}
\caption{Number of Major Airports per Continent and Country}
\label{fig:airport-count}
\end{figure}

We extracted flight data regarding 616 primary airports from 148 countries. Figure \ref{fig:airport-count} describes the number of major airports per continent and country. Currently, approximately  33\% of the world's airports belong in North America and 170 airports (27.6\%) in the United States. 27\% (166) and 26\% (162) belong to Asia and Europe respectively, and China (5.7\%) and the United Kingdom (4.4\%) have the highest number of airports within each continent.

\subsection{Number of Flights per Area and Country}
\begin{figure}[htb]
\centerline{\includegraphics[width=\hsize]{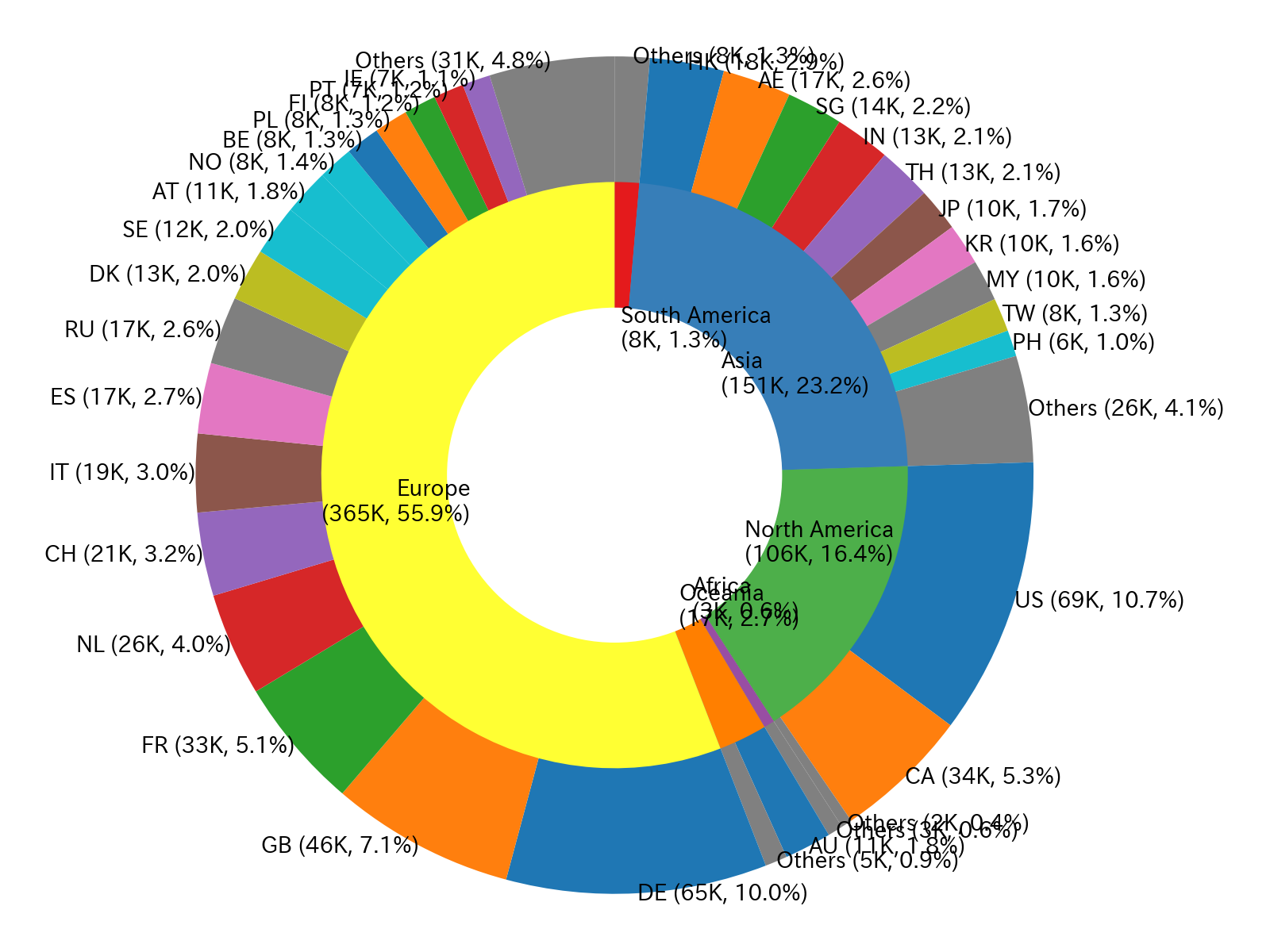}}
\caption{Number of Departed Flights per Continent and Country}
\label{fig:flights-count}
\end{figure}

Figure \ref{fig:flights-count} describes the total number of flights from January 1 to April 30 2020 for each continent and country. More than half of flights in this dataset departed from Europe. 10\% of international flights departed from Germany (10\%) and the United States (10.7\%). In Asia, Hong Kong, the United Arab Emirates, and Singapore had the highest number of flights (2.9\%, 2.6\%, and 2.2\% respectively), yet there were very few flights connecting to China. The number of flights in South America, Africa, and Oceania were too small to analyze, therefore these continents were excluded from the following sections.

\subsection{Number of Potential Passengers}
The OpenSky Network Dataset does not contain the final number of passengers for each flight. However, since the data contained each aircraft type used, we used this to estimate the number of passengers for each trip. Since the information regarding the flights were extracted from the capacity data of each aircraft type from their respective Wikipedia pages \cite{aircraft-passengers}, we aim to obtain more accurate information on aircraft types and their capabilities for our future analysis.

The proportion of international passengers travelling from Asia was approximately 31.2\% of all passengers, and the number of flights were 23.5\% of overall flights. On the other hand, the total number of passengers of international flights within Europe is about 46.5\% while the proportion of international flights outside of Europe is 55.5\%. The gap of proportions between passengers and flights comes from the difference in capacities of the aircraft.

\section{Descriptive Analysis of Time-series Flight Data}
\label{sec:timeseries}

\subsection{Overview}
Announcements and enforcement regarding lockdowns (border closures and travel restrictions, etc.) were implemented to slow the rate of transmissions and prevent overburdening of health facilities. The date of implementation, length, and extent vary by country. In the US, the Secretary of Health and Human Services declared a public health emergency on January 31\cite{national-emergency}. The US government suspended incoming flights from Europe (Schengen Area) on March 14\cite{us-close-eu}, and closed the border between the US and Canada on March 18\cite{us-close-canada}. Germany closed borders with neighboring countries on March 16\cite{germany-border}. Countries in the European Union agreed to reinforce border closures and restrict non-essential travels from March 17\cite{eu-close} onwards. In this section, we show the changes in the number of daily flights in each region, and discuss the correlations between travel restrictions and the number of flights in each country.

\subsection{Overall International Flights}
\begin{figure}[htb]
\centerline{\includegraphics[width=\hsize]{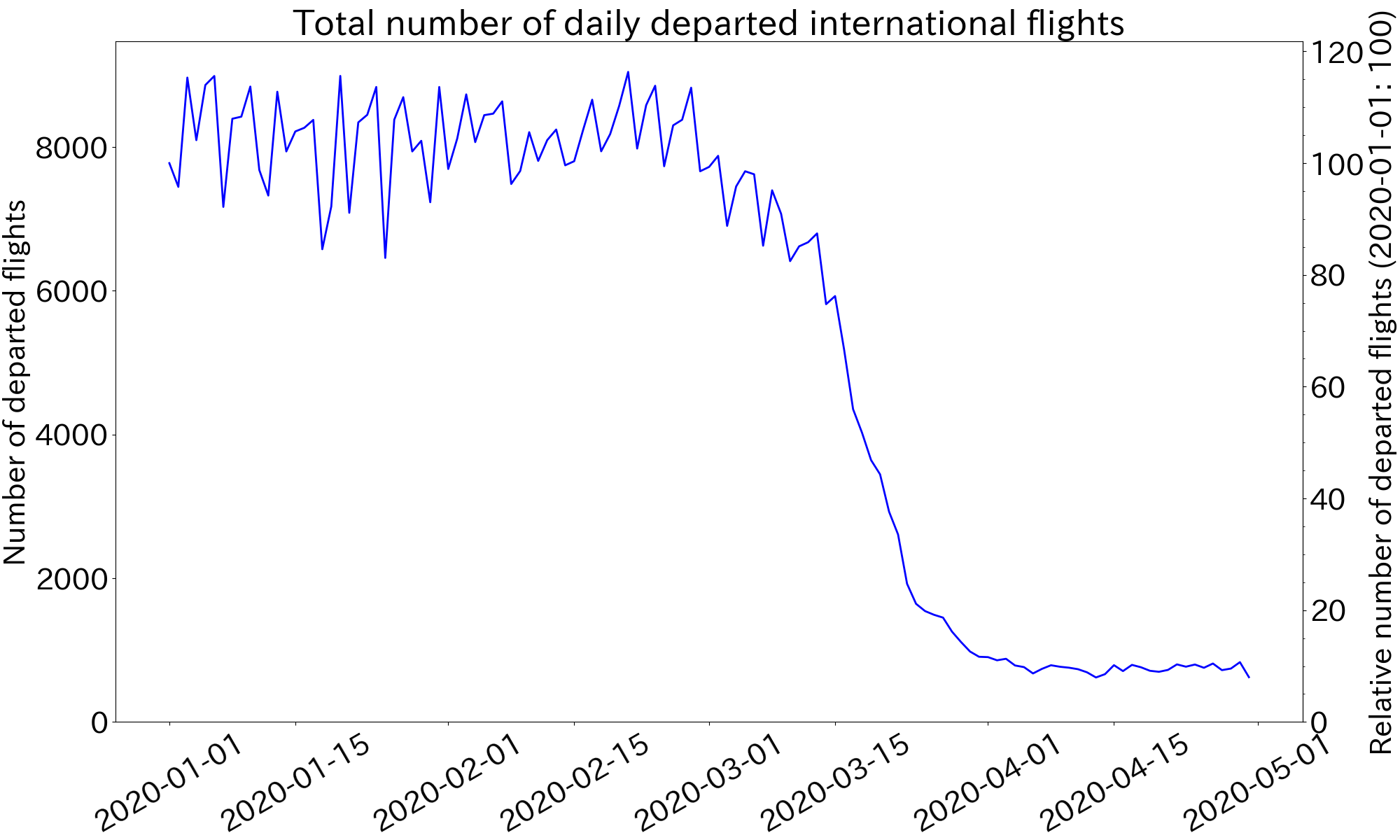}}
\caption{Total Number of Daily Departed International Flights}
\label{fig:global-intl}
\end{figure}

Figure \ref{fig:global-intl} describes the number of daily international flights in the world. Before the end of February, the daily number of international flights was approximately 8,000. However, the number of daily flights gradually decreased to 6,000 by early March and then drastically dropped to less than 1,000 (around 10\% of regular seasons) by the end of March.

\subsection{Inter-continental Flights}
\begin{figure}[htb]
\centerline{\includegraphics[width=\hsize]{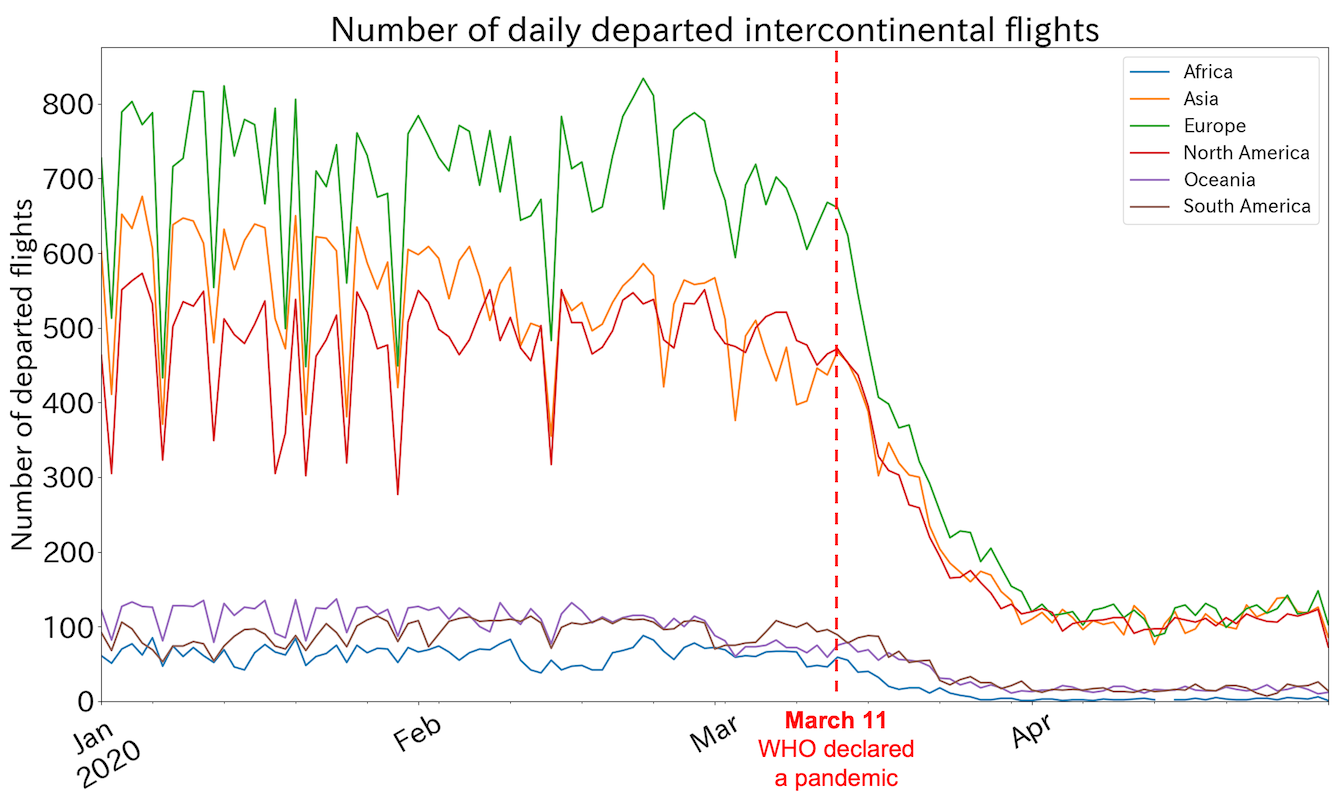}}
\caption{Total Number of Daily Departed Inter-continental Flights per Continent}
\label{fig:continent-intl}
\end{figure}

Figure \ref{fig:continent-intl} describes the number of intercontinental flights that departed from each continent. Inter-continental flights from Europe, Asia, and North America drastically decreased from March 13 as many flights were canceled. The decrease probably reflects the declaration of pandemic by WHO on March 11 and national emergency in the United States on March 13.

\subsection{Continental Flights}
\begin{figure}[htb]
\centerline{\includegraphics[width=\hsize]{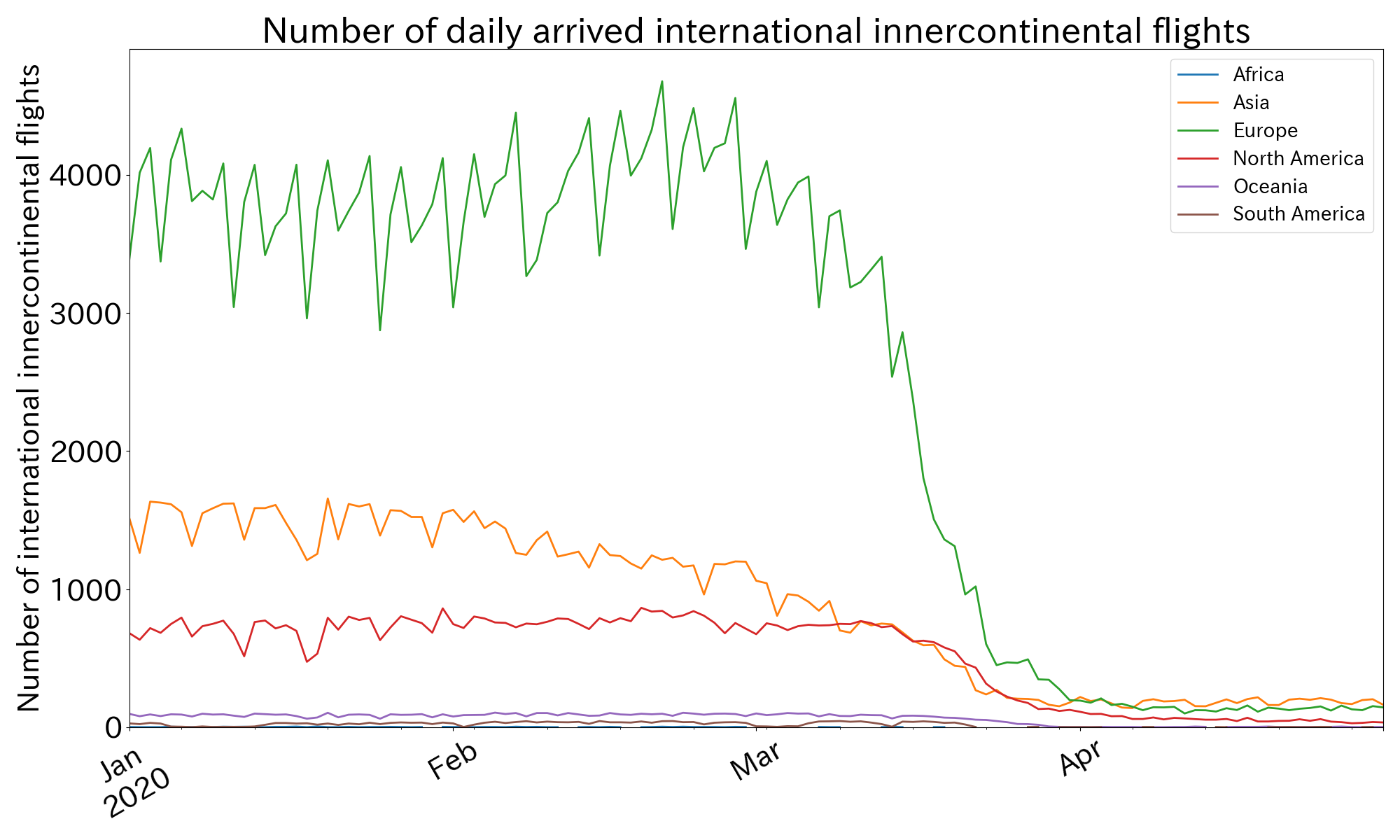}}
\caption{Number of Daily International Flights per Continent}
\label{fig:continent-inner}
\end{figure}

Figure \ref{fig:continent-inner} describes the number of continental flights within each continent.
In Europe, more than 3,000 flights were operating per day until the beginning of March, but dropped to 10\% of the regular number of flights from mid March to the end of March.
In Asia, the number of flights has gradually decreased from February.
In North America, the number of flights was steady until mid-March, before dropping significantly.

\begin{figure}[htb]
\centerline{\includegraphics[width=\hsize]{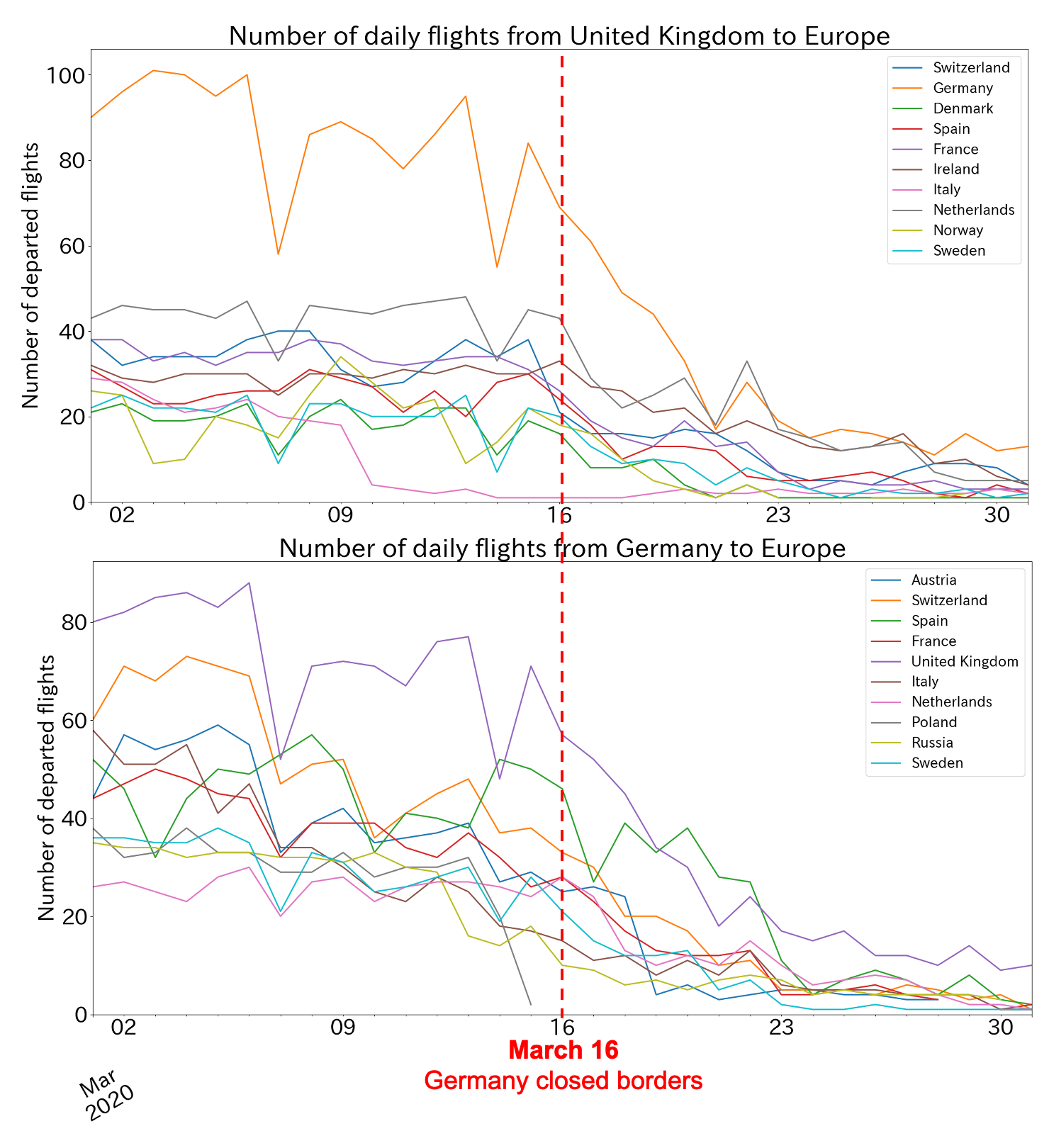}}
\caption{Number of Daily Flights from the United Kingdom and Germany to Other European Countries}
\label{fig:uk-de}
\end{figure}

In Europe, the UK, Ireland, and Germany had the highest number of flight passengers\cite{wb-passengers}. In Germany, international flights decreased from the beginning of March and then dropped after the nation closed its borders on March 16 (Figure \ref{fig:uk-de}). This reflected the EU's agreement to ban incoming travel from areas outside the EU, UK, and Switzerland on March 17.

\begin{figure}[htb]
\centerline{\includegraphics[width=\hsize]{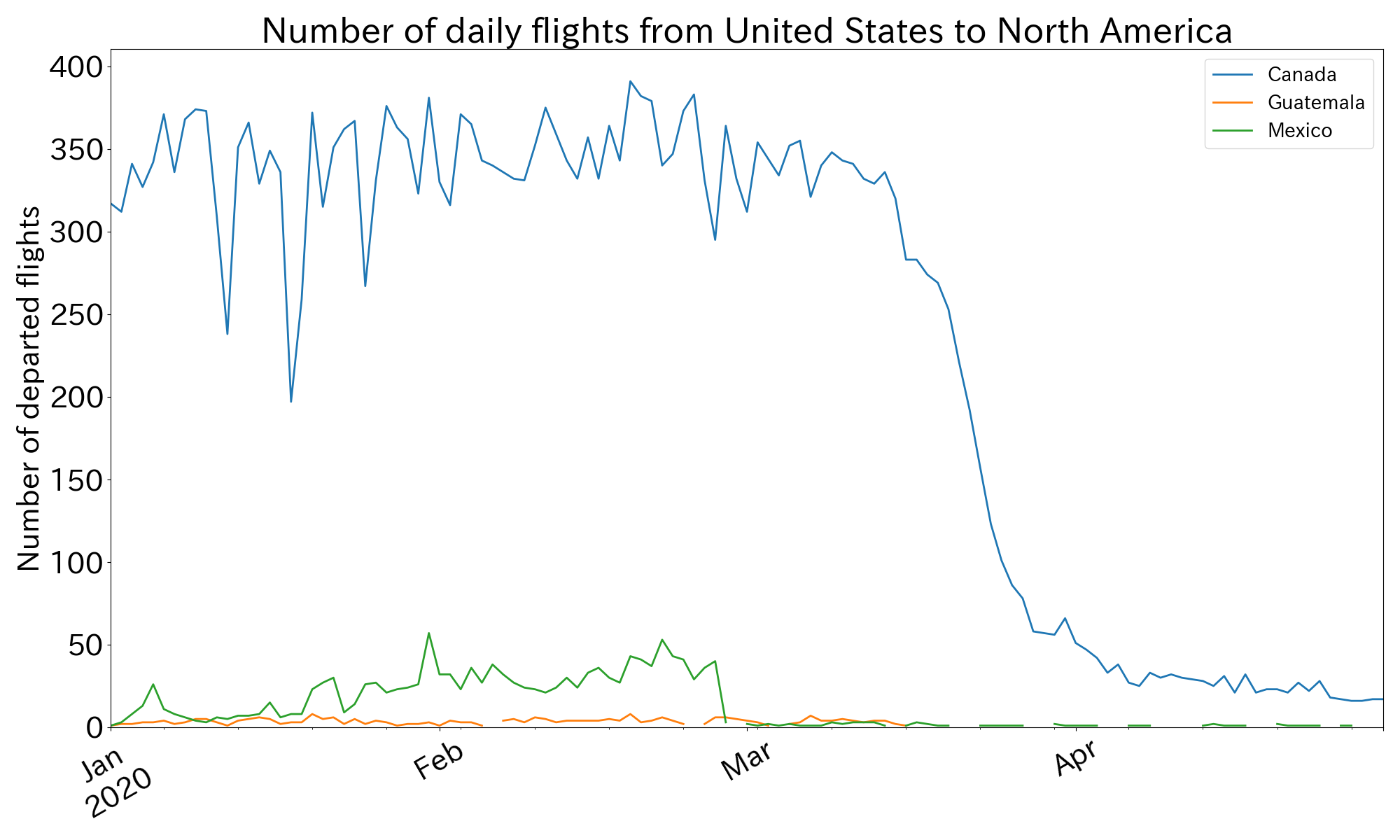}}
\caption{Number of Daily International Flights from the United States in North America}
\label{fig:us-na}
\end{figure}

Most of the continental flights had been operating in North America (the US in particular) until mid-March and half number of these flights in the Beginning of January are still in operation today (Figure \ref{fig:us-na}). On the other hand, the number of flights in Asia, Europe, and Oceania have dropped to less than 20\% to that of January 1.

\subsection{Domestic and International Flights in the United States}
\begin{figure}[htb]
\centerline{\includegraphics[width=\hsize]{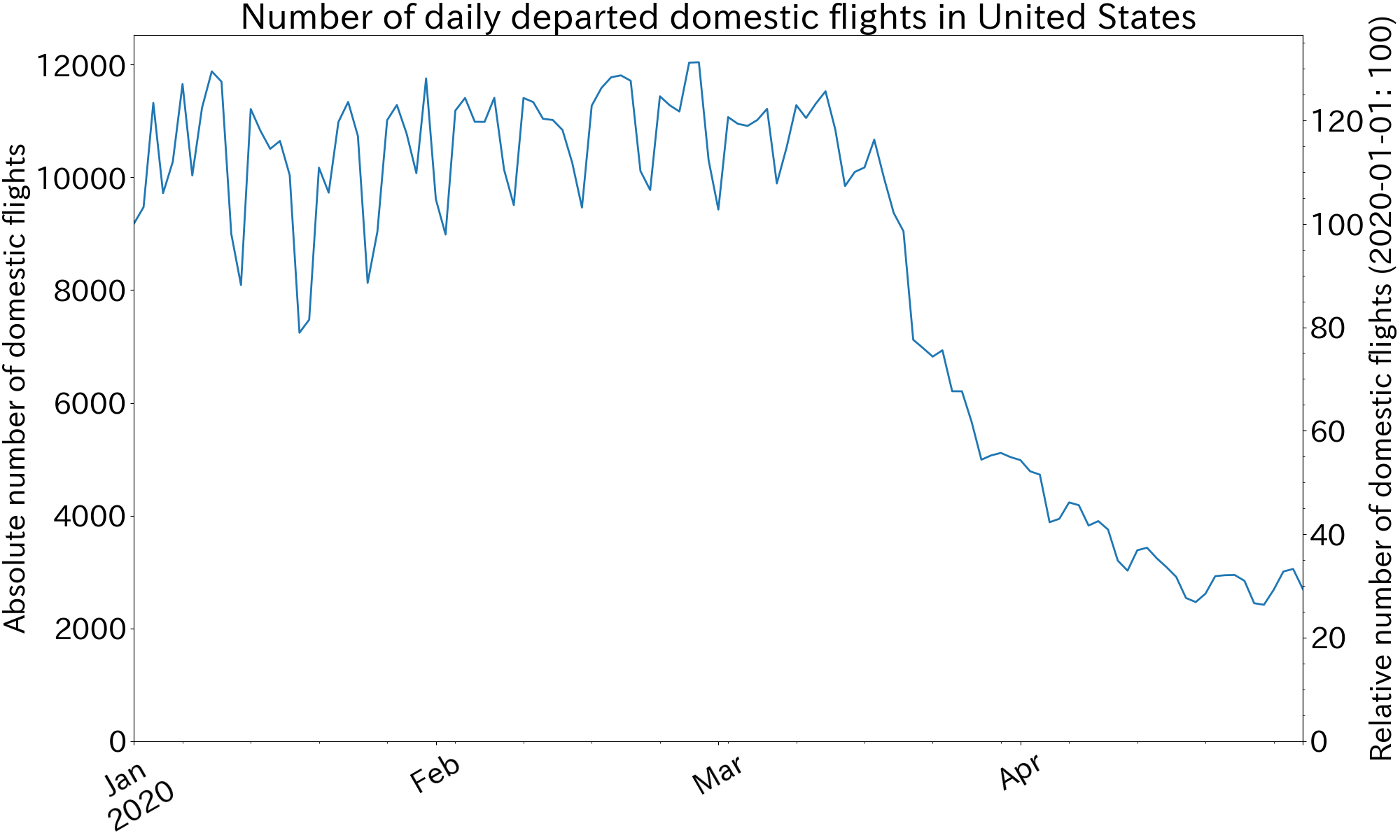}}
\caption{Number of Domestic Flights in the United States}
\label{fig:us-domestic}
\end{figure}

In the US, the number of domestic flights suddenly dropped from mid-March. However, roughly 50\% of all regular flights were still in operation on March 31 (Figure \ref{fig:us-domestic}), although the US government and airlines planned to ban or reduce flights by that point \cite{us-flights}. In April, the number of flights gradually declined but 30\% of regular flights were still in operation.

\begin{figure}[htb]
\centerline{\includegraphics[width=\hsize]{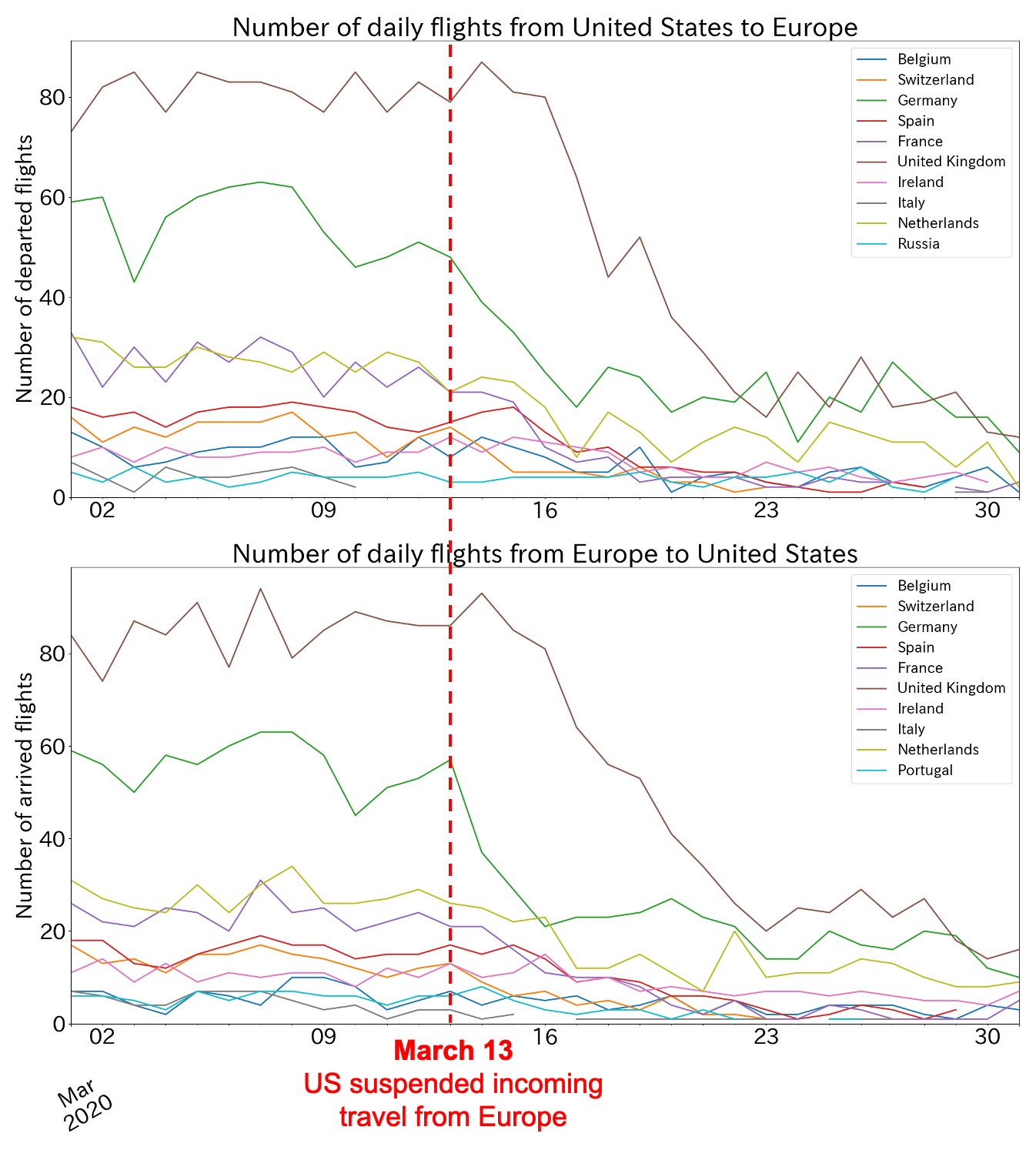}}
\caption{Number of Flights Between the United States and European Countries}
\label{fig:us-eu}
\end{figure}

\begin{figure}[htb]
\centerline{\includegraphics[width=\hsize]{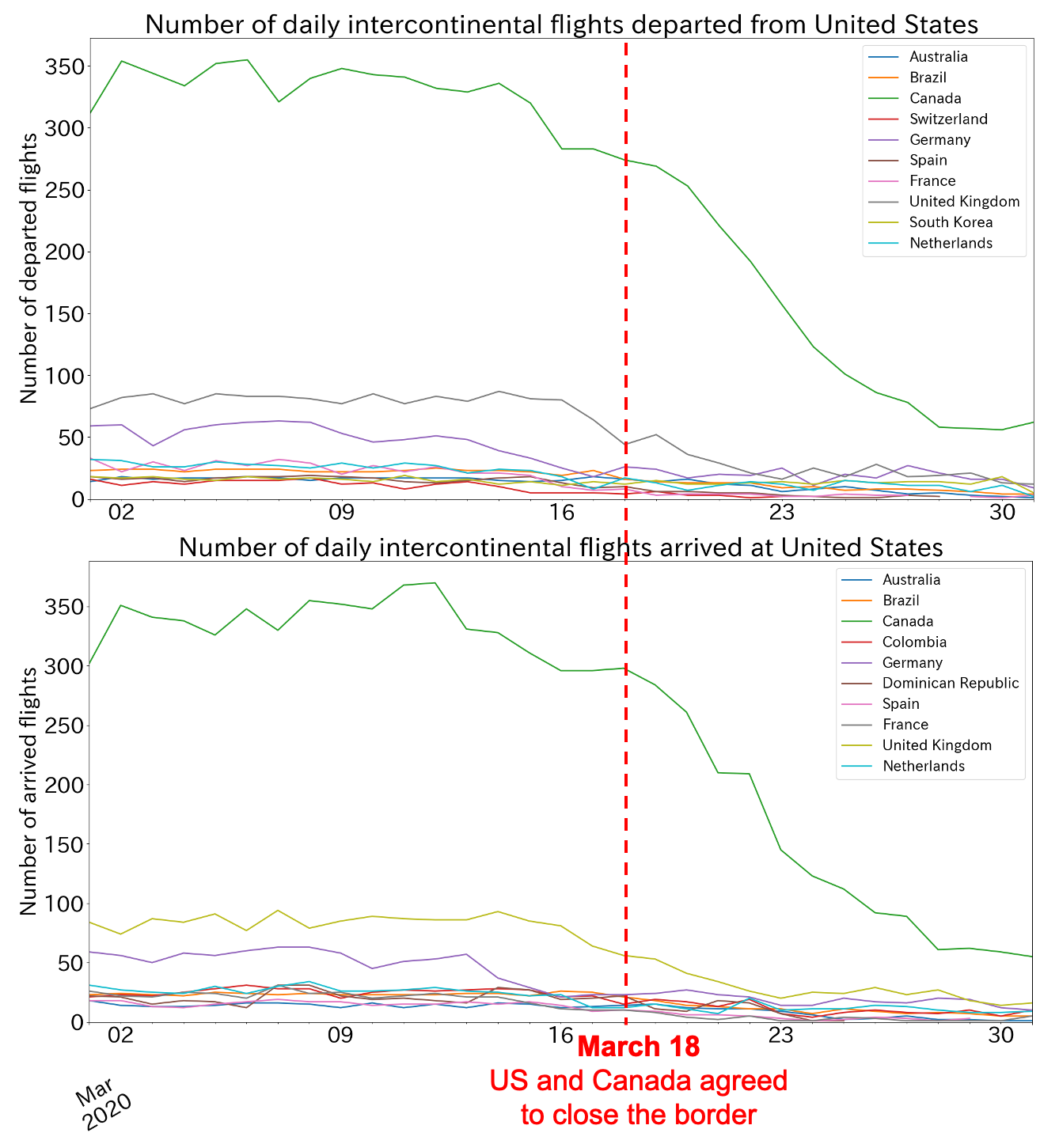}}
\caption{Number of Flights Between the United States and top-10 Countries}
\label{fig:us-top10}
\end{figure}

The number of flights from European countries to the US suddenly dropped after March 14 (Figure \ref{fig:us-eu}). The US suspended incoming travel from Europe (Schengen area) from March 13. In the overall number of international flights, the primary destination from the US is Canada (green lines in Figure \ref{fig:us-top10}). The number of daily flights between the US and Canada gradually decreased after these governments agreed to close their border on March 18.

\subsection{Key Takeaways}
We analyzed the number of daily flights for each country and continent, and compared the relationship between the date of travel restrictions and the decline of flights. We found that many international flights suddenly declined during the middle of March. In Europe, the number of flights dropped to about 10\% of regular seasons after the EU agreed to restrict incoming travels on March 17. In the US, approximately half of all domestic flights were still in operation even at the end of March.

\section{Graph Analytics on the Flight Network}
\label{sec:graph}

\subsection{Overview}
With travel restrictions, the global flight network gradually became more sparse, which prevented travelers from:
\begin{enumerate}
    \item Using direct flights to their destinations
    \item Returning to their origin once they depart from the airport
    \item Departing from their current location due to absence of flights in nearby airports
\end{enumerate}

By constructing daily flight networks, we quantitatively evaluated these effects through graph analytics. Each vertex and edge represent an airport and a flight, respectively.
\begin{enumerate}
    \item Distribution of the number of neighboring airports and flights (degree and neighborhood distributions)
    \item Strongly connected component (SCC) extraction
    \item Isolated vertex extraction
\end{enumerate}

\begin{figure}[htb]
\centerline{\includegraphics[width=\hsize]{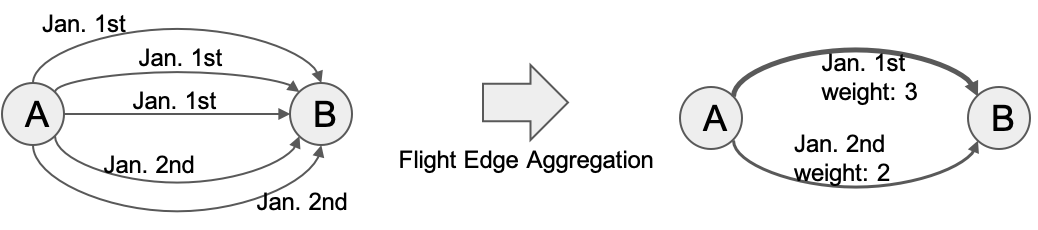}}
\caption{Flight Edge Aggregations by Date}
\label{fig:agg-edges-date}
\end{figure}

In the construction of flight networks, we aggregated multiple flights with the same origin and destination and date into a single weighted edge in the flight network construction like Figure \ref{fig:agg-edges-date}. The edge weight is the total number of flights.

\begin{figure}[htb]
\centerline{\includegraphics[width=\hsize]{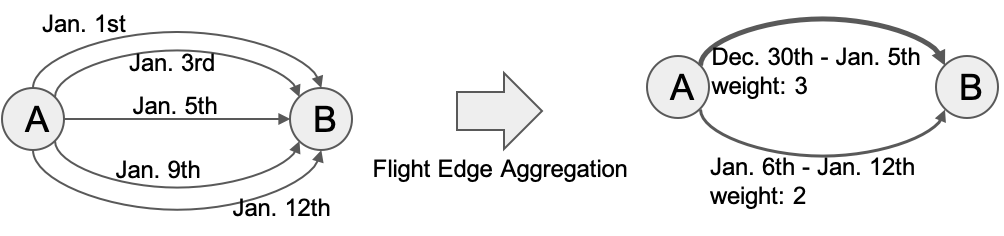}}
\caption{Flight Edge Aggregations by Week}
\label{fig:agg-edges-week}
\end{figure}

While multiple flights are in operation every day, not all flights are operated every day (e.g., three days a week). To detect long-term trends more precisely, we also aggregated flight edges by week as an optional preprocessing (Figure \ref{fig:agg-edges-week}).

\subsection{Connecting Airports and Flights}
\begin{figure}[htb]
\centerline{\includegraphics[width=\hsize]{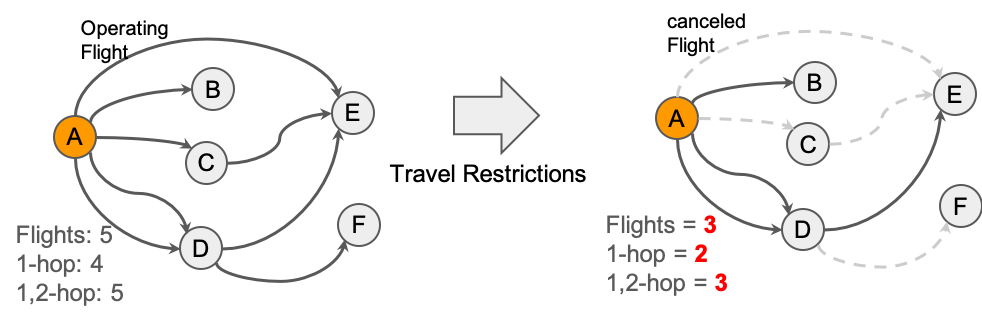}}
\caption{Number of Flights and Neighboring Airports from the Airport $A$}
\label{fig:degree}
\end{figure}

In regular times, many flights (edges in the flight network) connect from the airport (vertices in the flight network) to the airport. That helps travelers to get to more destinations with fewer transfers. However, travel restrictions prevent them from using direct flights and getting to their destinations. To measure the effect quantitatively, we computed the following metrics for each airport and each date (Figure \ref{fig:degree}).

\begin{enumerate}
    \item Number of flights (edges) from the airport
    \item Number of 1-hop neighboring airports from the airport (the number of reachable destinations with a nonstop flight)
    \item Number of 1,2-hop neighboring airports from the airport (the number of reachable destinations with a single connecting flight)
\end{enumerate}

\begin{figure}[htb]
\begin{tabular}{c}
\begin{minipage}{0.9\hsize}
    \centerline{\includegraphics[width=0.9\hsize]{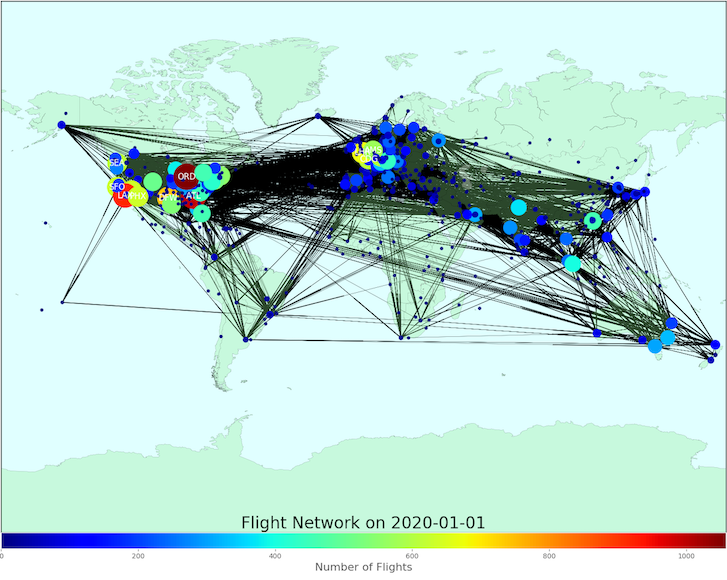}}
    \caption{Global Flight Network on January 1}
    \label{fig:global-map-jan}
\end{minipage}\\
\begin{minipage}{0.9\hsize}
\centerline{\includegraphics[width=0.9\hsize]{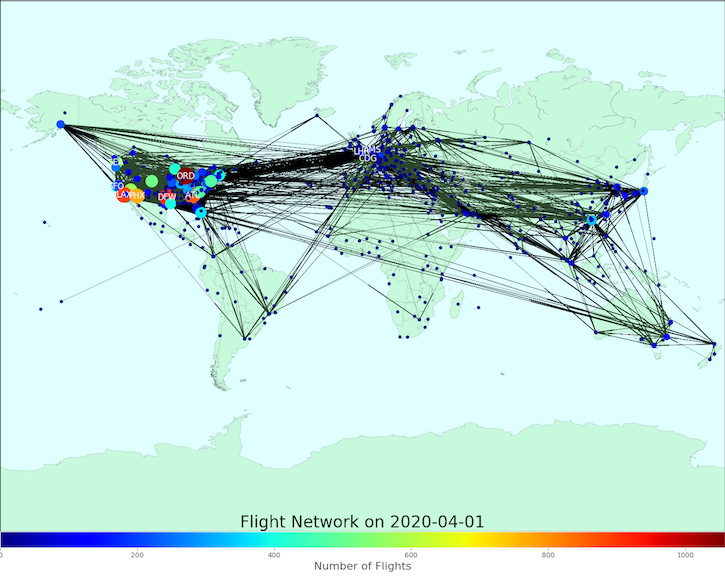}}
    \caption{Global Flight Network on April 1}
    \label{fig:global-map-apr}
\end{minipage}
\end{tabular}
\end{figure}

\begin{figure}[htb]
\begin{tabular}{c}
\begin{minipage}{0.9\hsize}
    \centerline{\includegraphics[width=0.9\hsize]{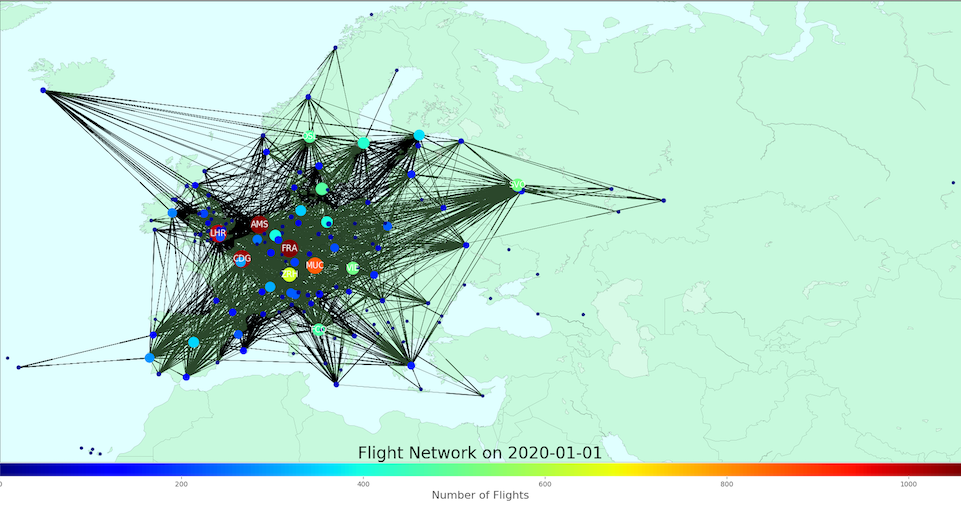}}
    \caption{Flight Network in Europe on January 1}
    \label{fig:eu-map-jan}
\end{minipage}\\
\begin{minipage}{0.9\hsize}
\centerline{\includegraphics[width=0.9\hsize]{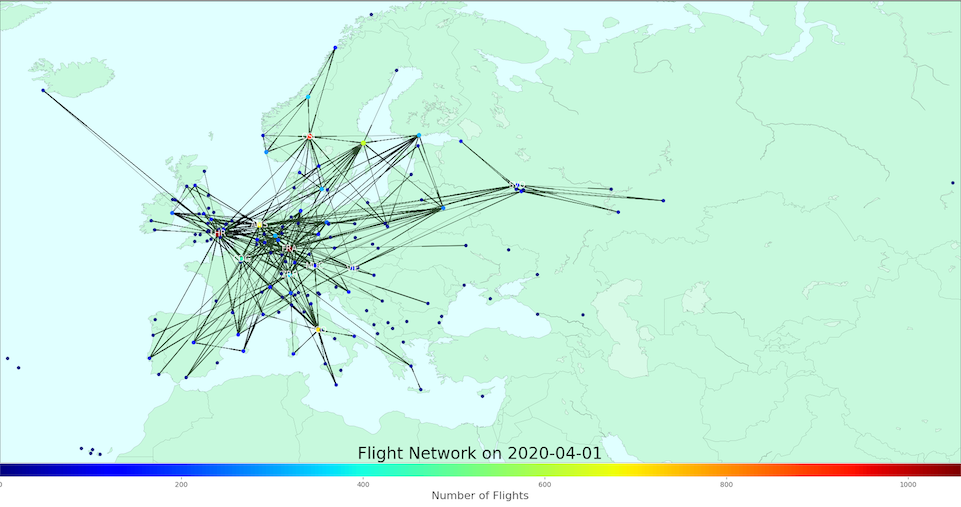}}
    \caption{Flight Network in Europe on April 1}
    \label{fig:eu-map-apr}
\end{minipage}
\end{tabular}
\end{figure}

\begin{figure}[htb]
\begin{tabular}{c}
\begin{minipage}{0.9\hsize}
    \centerline{\includegraphics[width=\hsize]{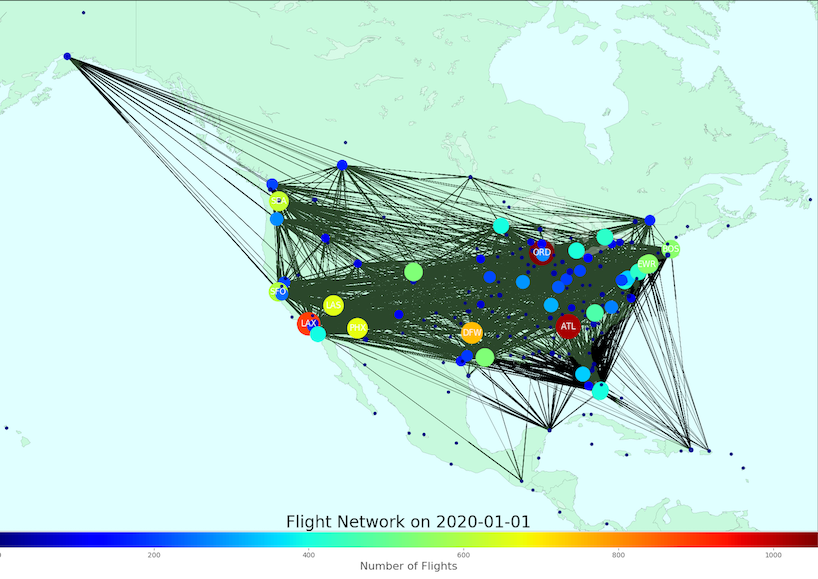}}
    \caption{Flight Network in North America on January 1}
    \label{fig:na-map-jan}
\end{minipage}\\
\begin{minipage}{0.9\hsize}
\centerline{\includegraphics[width=\hsize]{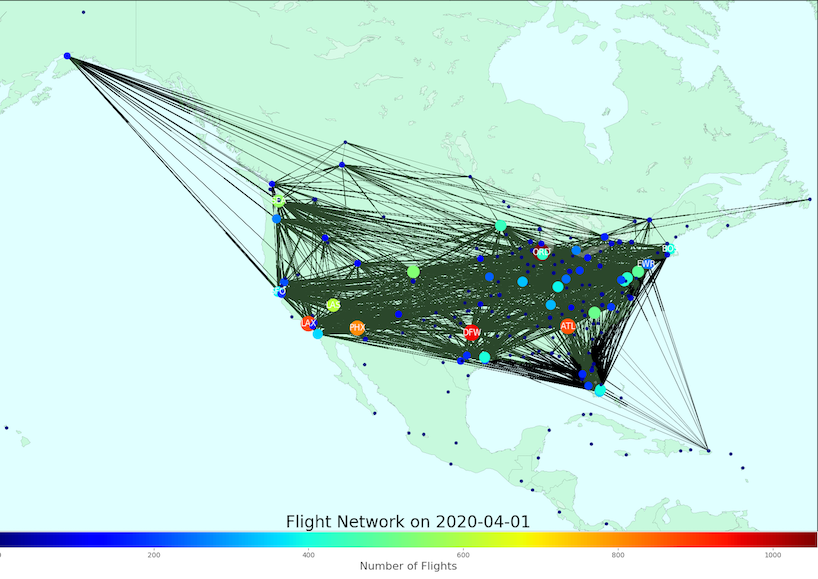}}
    \caption{Flight Network in North America on April 1}
    \label{fig:na-map-apr}
\end{minipage}
\end{tabular}
\end{figure}

The color and size of each vertex in Figure \ref{fig:global-map-jan} - \ref{fig:na-map-apr} indicate the number of departed flights from the airport. From mid-March to the beginning of April, many flights were gradually canceled, except for domestic flights in the United States and international flights between the United States and Europe.

The number of flights in Europe suddenly decreased from March 16 (the day before EU countries closed borders) to March 23. In April, only a few flights were operating and many airports became isolated.

In North America, most of the flights in North America were operating between the United States and Canada. In the United States, most of the domestic flights were operating as usual, even at the end of April.

\subsection{Density and Clustering Coefficient}
\label{sec:density}

With denser flight networks, travelers can reach more destinations. As the barometer of the flight network density, we computed the following metrics.\\
\textbf{Density:} The ratio of the actual number of edges after aggregation against the number of edges of the complete graph.\\
\textbf{Clustering Coefficient:} The ratio of the total number of triangles against the total number of triplets (connected three vertices). This is mainly used for social network analytics to measure the connectivity of friendships.

\begin{figure}[htb]
\centerline{\includegraphics[width=\hsize]{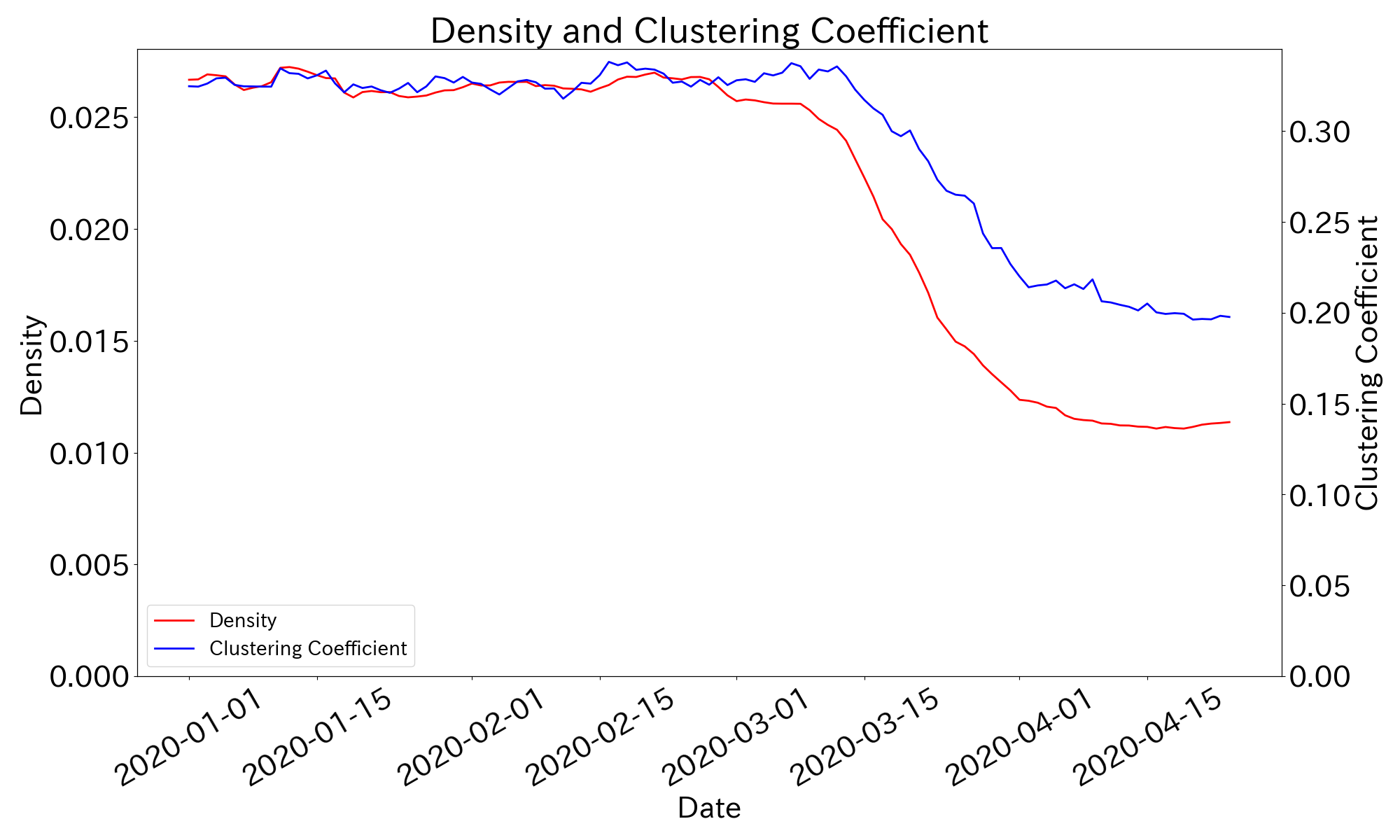}}
\caption{Density (red line) and Clustering Coefficient (blue line) of the Global Flight Network}
\label{fig:global-density}
\end{figure}

\begin{figure}[htb]
\centerline{\includegraphics[width=\hsize]{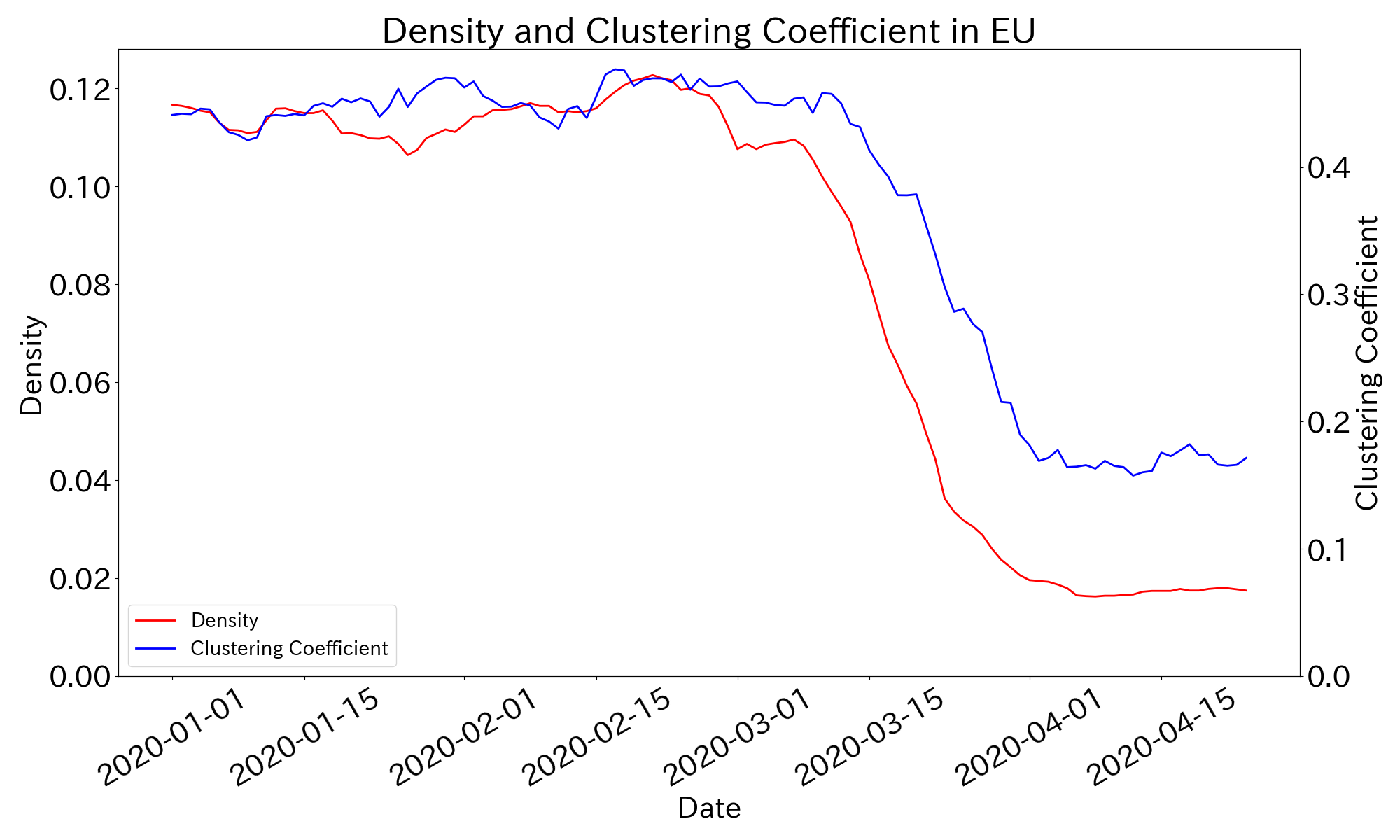}}
\caption{Density (red line) and Clustering Coefficient (blue line) of the Flight Network in Europe}
\label{fig:eu-density}
\end{figure}

\begin{figure}[htb]
\centerline{\includegraphics[width=\hsize]{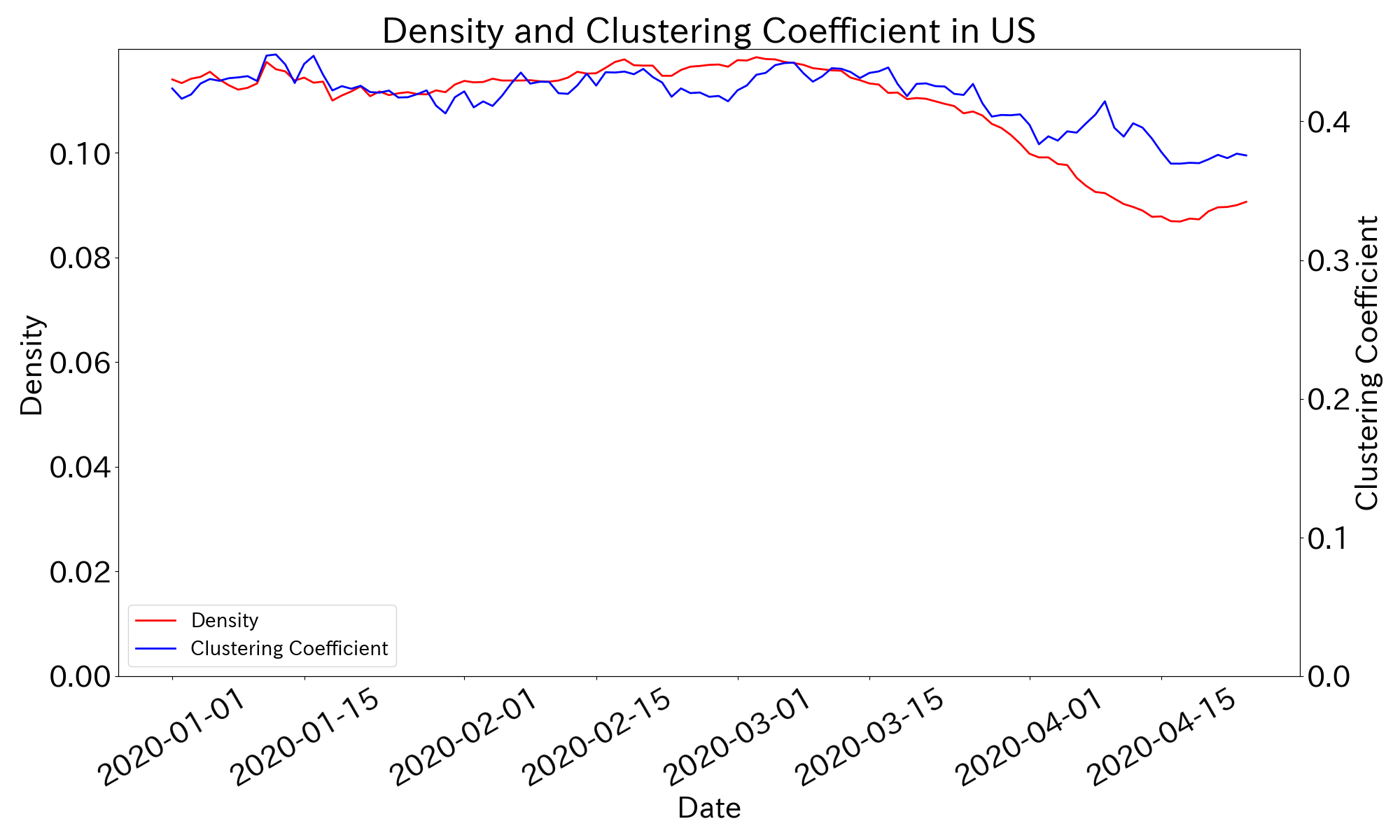}}
\caption{Density (red line) and Clustering Coefficient (blue line) of the Flight Network in the United States}
\label{fig:us-density}
\end{figure}

The density (red lines in Figure \ref{fig:global-density} - \ref{fig:us-density}) of the global flight network rapidly decreased by half during March 8 to April 1. In Europe, drastic decreases in the flight network density was similar to that of the global network, and fell to approximately 17\% of the peak. In the US on the other hand, the density gradually decreased, yet only by 20\%.
From March 13, the clustering coefficient (blue lines) of the global flight network gradually decreased by half in the next two weeks. In Europe, it rapidly declined to about a quarter in the same period. In the US, the decline of the clustering coefficient is only about 10\% at the end of April.

\subsection{Strongly Connected Component and Isolated Airports}
\begin{figure}[htb]
\centerline{\includegraphics[width=0.6\hsize]{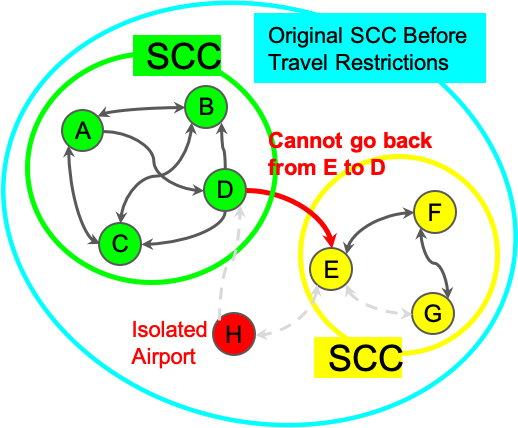}}
\caption{An Example of SCC Decompositions of Flight Networks}
\label{fig:scc-example}
\end{figure}

In regular seasons, travelers departing from any airport can typically go to any destination and return to their airport of origin. This is due to flight networks being a strongly connected component (SCC), within which travelers can go from/to any airports in the same SCC.
With travel restrictions, however, many flights were canceled and the large SCC (blue circle in Figure \ref{fig:scc-example}) has decomposed into smaller SCCs (yellow and green circles in Figure \ref{fig:scc-example}). In the result, travelers cannot:
\begin{enumerate}
    \item Go to other airports outside local flight networks
    \item Return to their origin (a red edge from E to D)
    \item Depart from isolated airports (a red vertex)
\end{enumerate}

In the global flight network, the size of the largest SCC (red line) decreased by 30\% from March 13 (300) to April 6 (210), and the number of isolated airports (blue line) increased by 30\% (from 310 to 400). In Europe, the largest SCC size on April 7 was less than half of March 10 (before border closures), and the number of isolated airports tripled. In the US on the other hand, the largest SCC size and isolated airports are almost stable (80 and 100 respectively). Although the largest SCC size gradually decreased from March 13 to April 6, the rate of the overall decrease is only about 10\%.

\begin{figure}[htb]
\centerline{\includegraphics[width=\hsize]{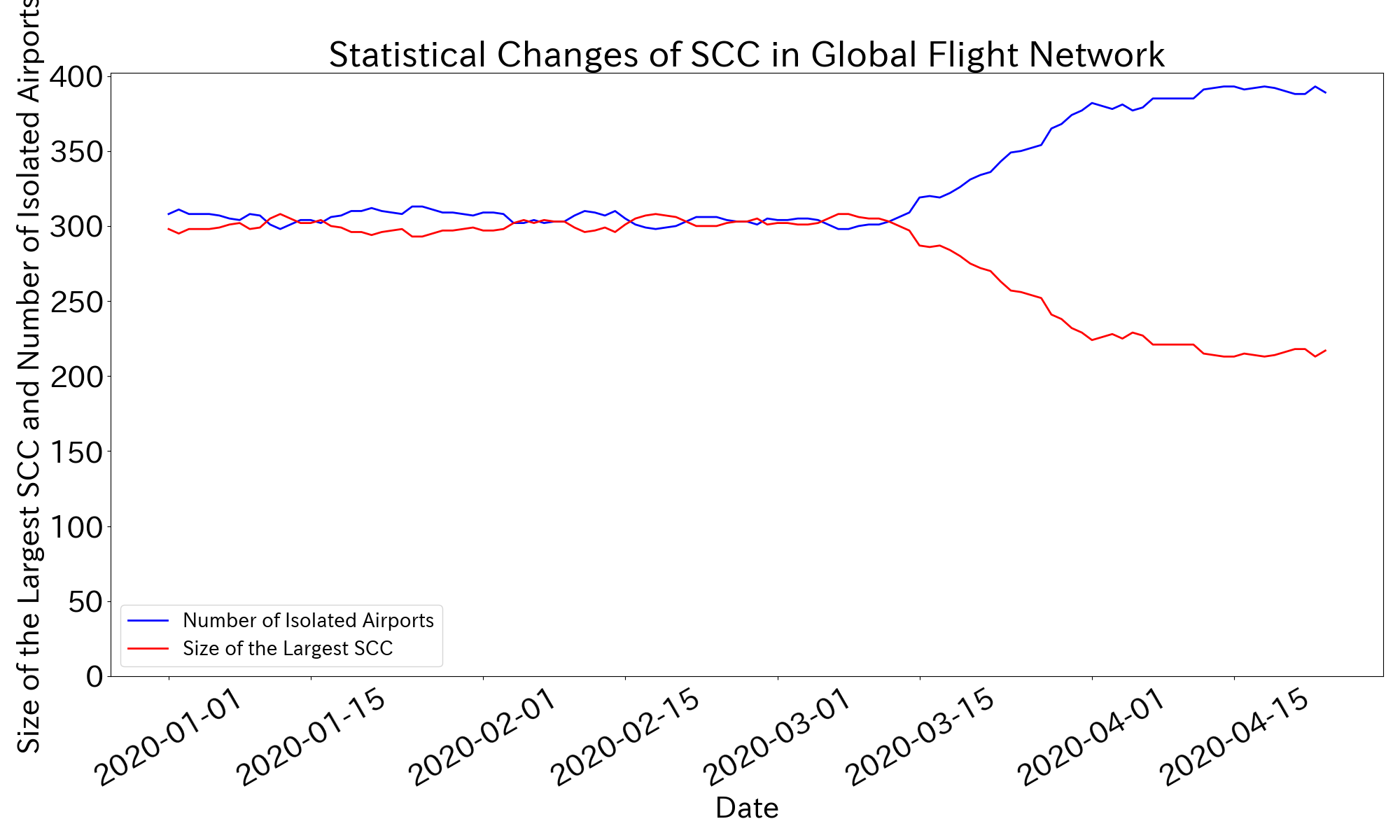}}
\caption{Largest SCC Size (red line) and the Number of Isolated Airports (blue line) of the Global Flight Network}
\label{fig:scc-global}
\end{figure}

\begin{figure}[htb]
\centerline{\includegraphics[width=\hsize]{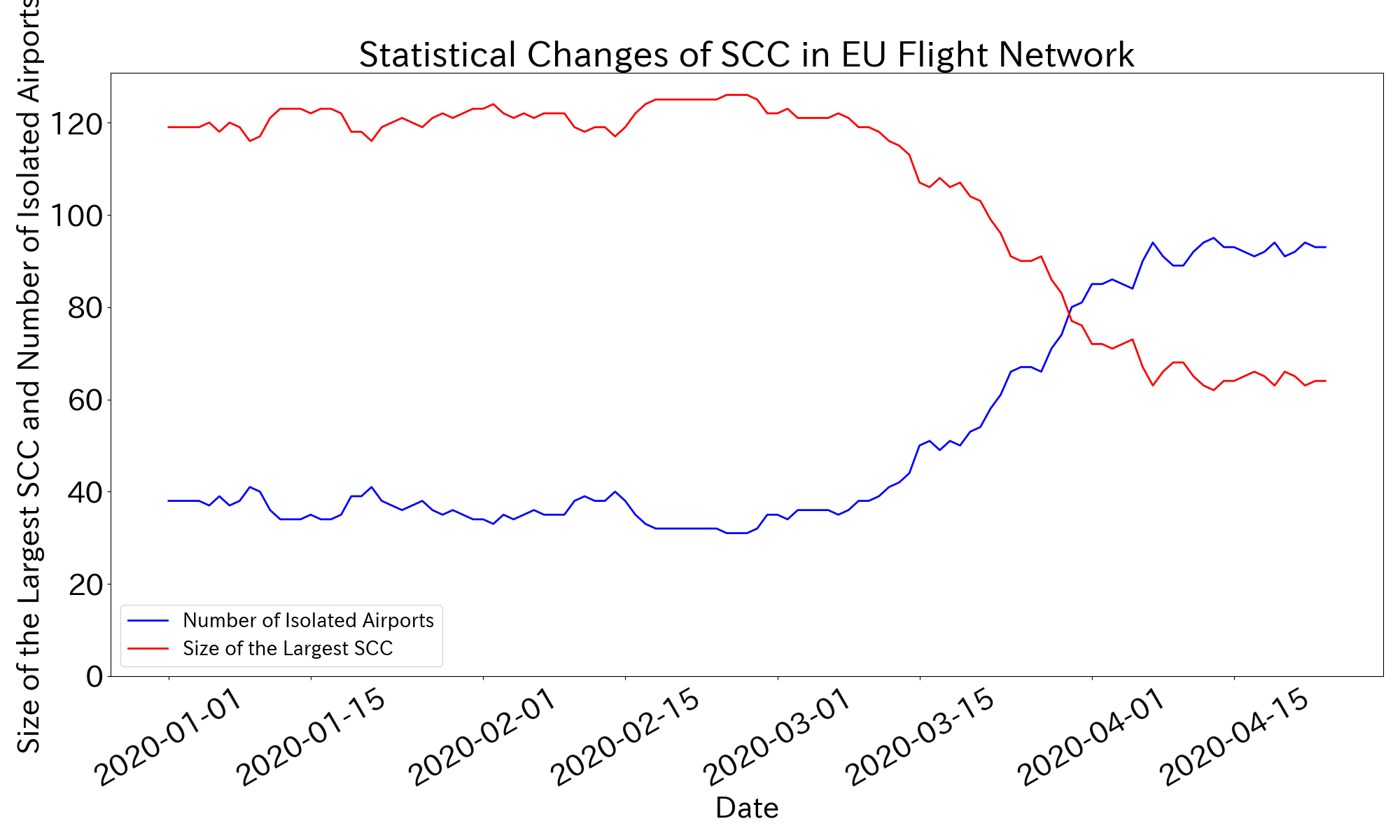}}
\caption{Largest SCC Size (red line) and the Number of Isolated Airports (blue line) of the Flight Network in Europe}
\label{fig:scc-eu}
\end{figure}

\begin{figure}[htb]
\centerline{\includegraphics[width=\hsize]{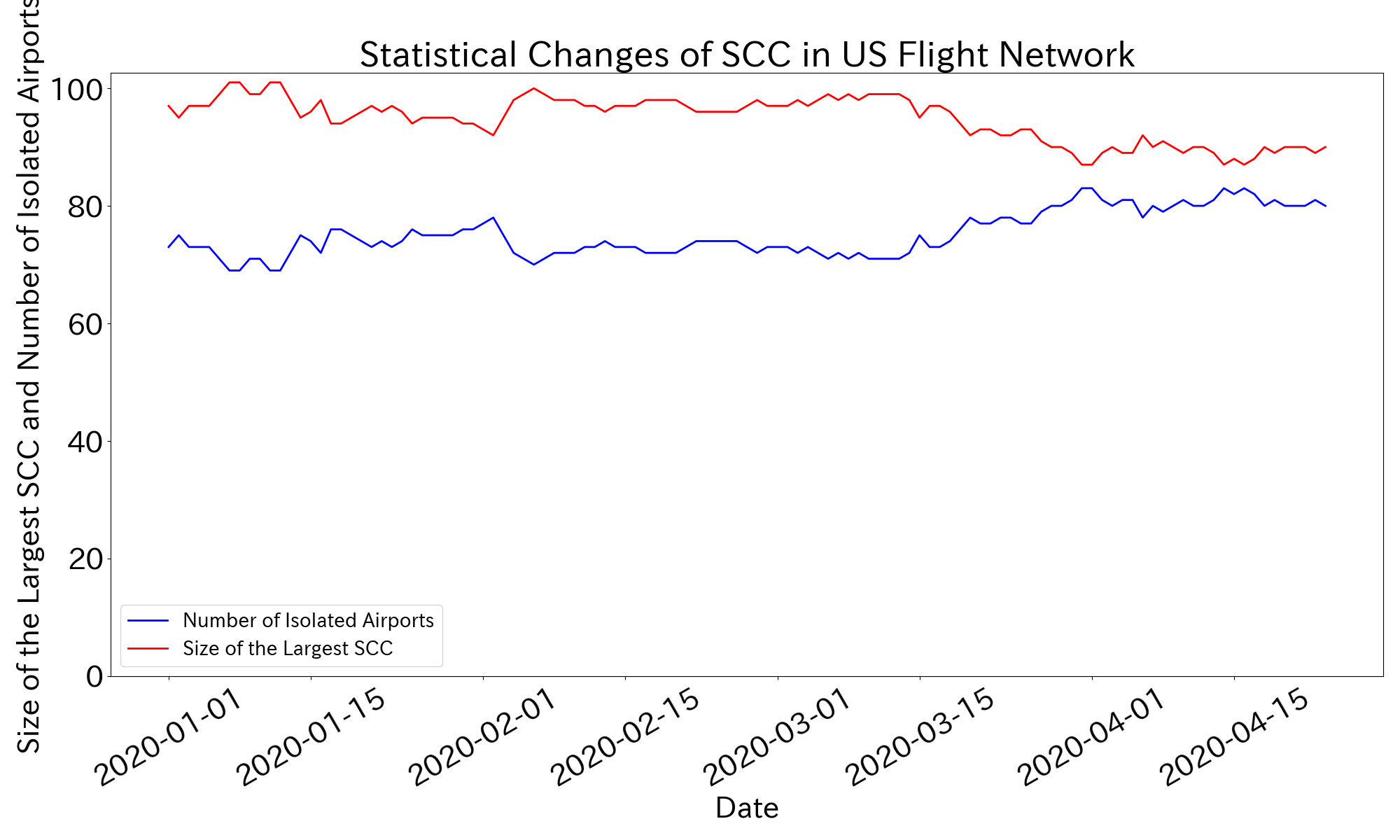}}
\caption{Largest SCC Size (red line) and the Number of Isolated Airports (blue line) of the Flight Network in the United States}
\label{fig:scc-us}
\end{figure}

\subsection{Diameter and Radius of Flight Network}
When many flights are canceled, more transfers are required for travellers to reach the destinations compared to regular seasons. In order to measure the number of required transfers, we applied a graph analysis to compute eccentricity (the maximum distance from a specified vertex to all other connected vertices) of the flight network. An eccentricity of an airport vertex corresponds to the number of required transfers to other destinations in the worst case. With travel restrictions, the diameter (the maximum value of eccentricities) and the radius (the minimum value of eccentricities) of the flight network become larger.

\begin{figure}[htb]
\centerline{\includegraphics[width=\hsize]{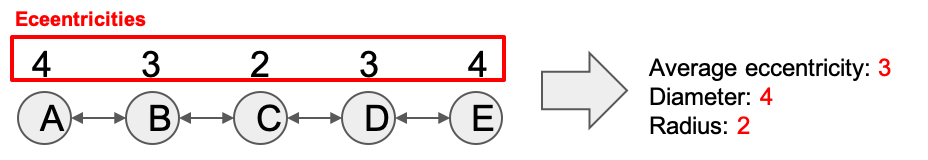}}
\caption{An Example of the Average Eccentricity, Diameter and Radius}
\label{fig:diaeter-example}
\end{figure}

\begin{figure}[htb]
\centerline{\includegraphics[width=\hsize]{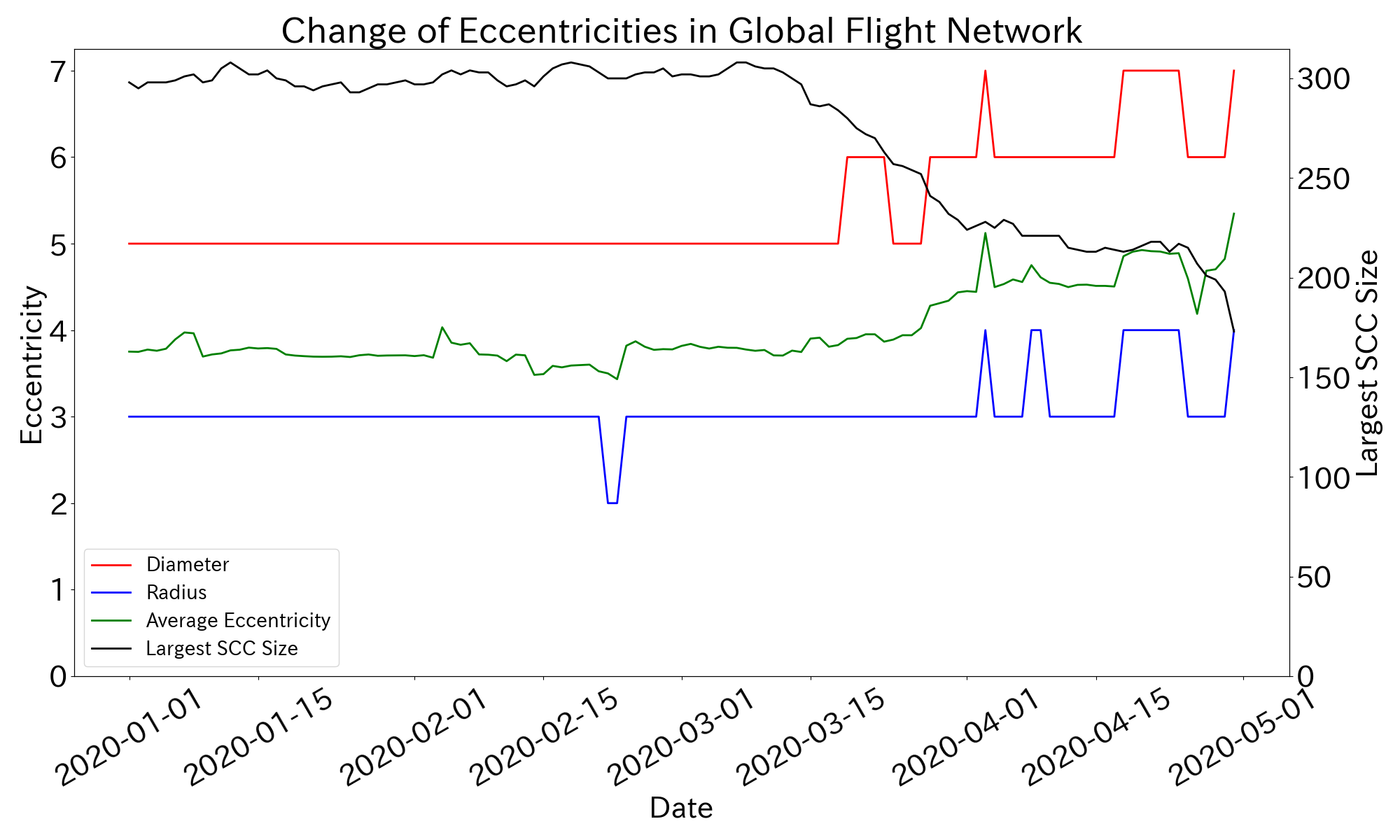}}
\caption{Diameter (red line) and Radius (blue line) of the Global Flight Network}
\label{fig:diaeter-global}
\end{figure}

\begin{figure}[htb]
\centerline{\includegraphics[width=\hsize]{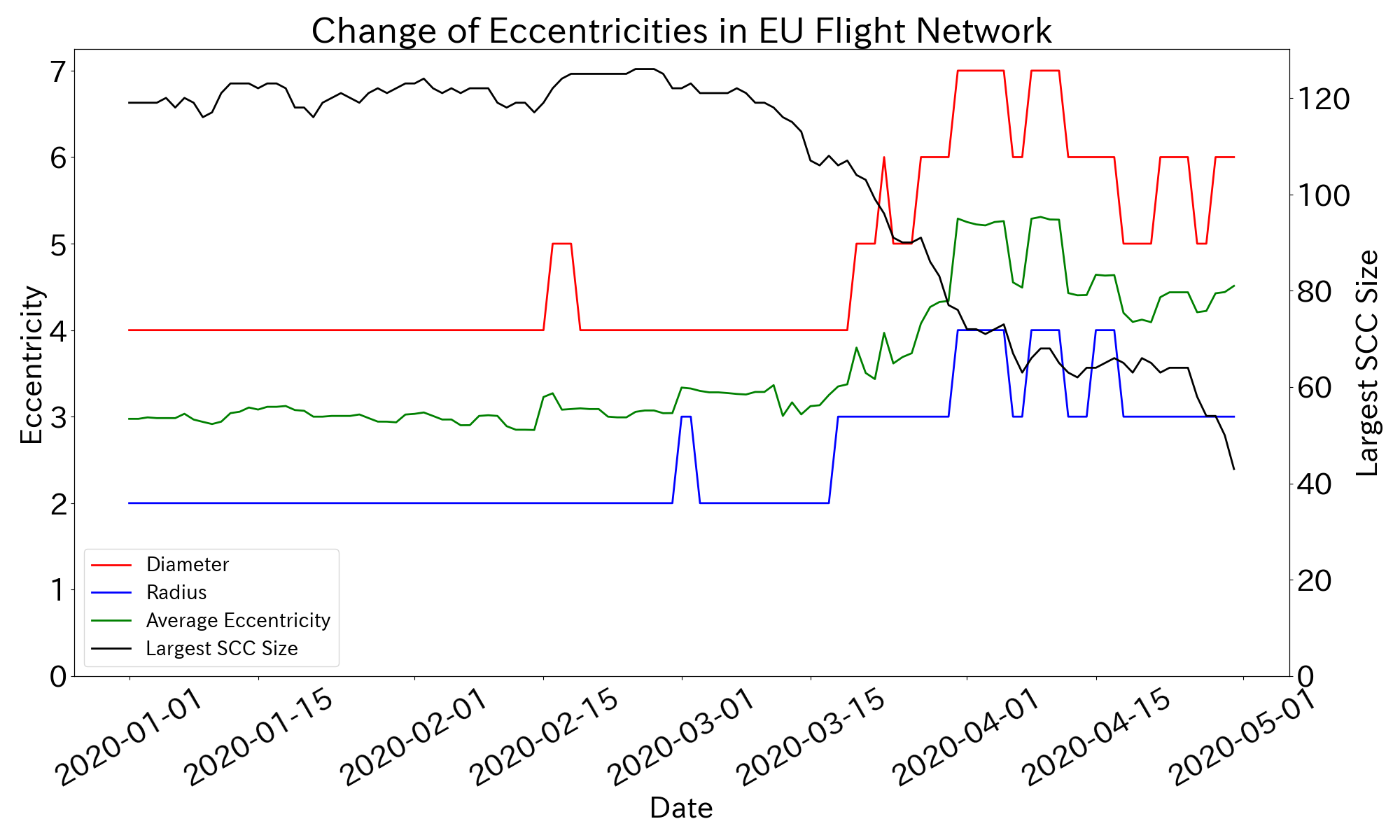}}
\caption{Diameter (red line) and Radius (blue line) of the Europe Flight Network}
\label{fig:diaeter-eu}
\end{figure}

\begin{figure}[htb]
\centerline{\includegraphics[width=\hsize]{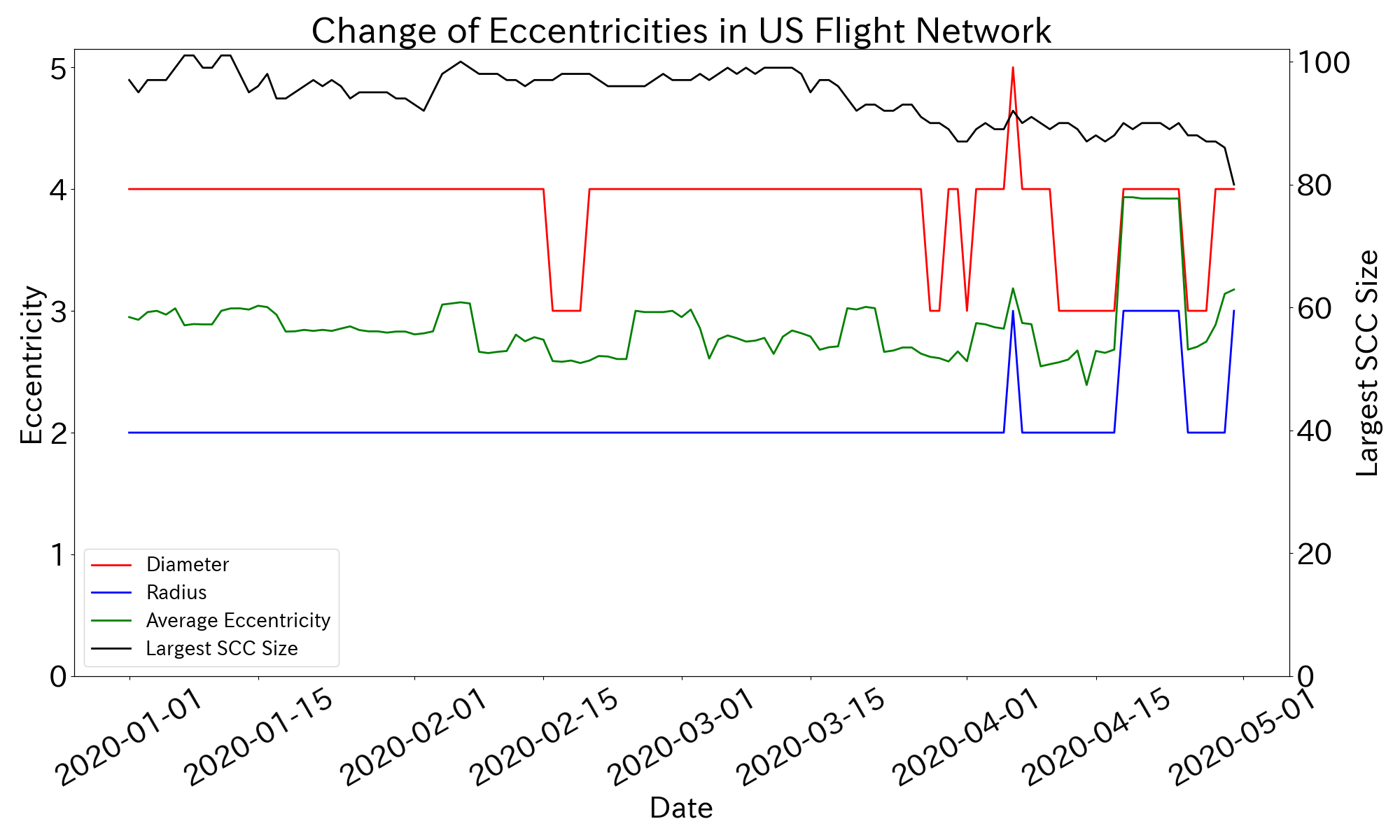}}
\caption{Diameter (red line) and Radius (blue line) of the United States Flight Network}
\label{fig:diaeter-us}
\end{figure}

The diameter (red line in Figure \ref{fig:diaeter-global} - \ref{fig:diaeter-us}), radius (blue line), and average eccentricity (green line) of the largest SCC size (black line) in the global flight network increased from April. In the flight network in Europe, diameter and radius rapidly increased from March 16 in two weeks. The overall flight network’s trend in the US did not change until the end of April.

\subsection{Flight Edge Weight and Density}
Although a number of flights still exist among most of the airports, the flight frequency has gradually decreased. To measure the decline of flight frequencies, we defined the “relative weight” for each flight edge (pair of vertices as connected airports) and network density with the weighted edges.

First, we computed the moving average of the number of daily flights for each pair of vertices, and then computed the ratio of the total number of flights in a week against the baseline of the specified week (January 12 - January 18), before defining the ratio as the relative weight of the edge. Besides the topological density described in \ref{sec:density}, we computed the density of the whole flight network with the weighted edges.

\begin{figure}[htb]
\centerline{\includegraphics[width=\hsize]{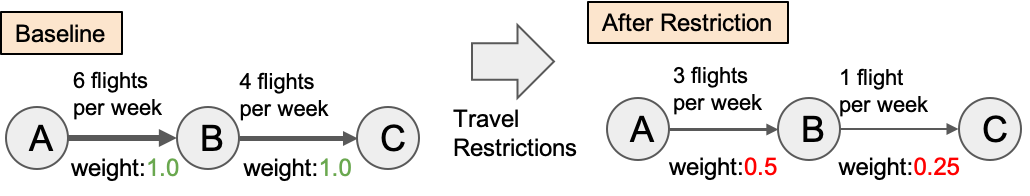}}
\caption{An Example of Weighted Flight Edges}
\label{fig:weight-example}
\end{figure}

\begin{figure}[htb]
\centerline{\includegraphics[width=\hsize]{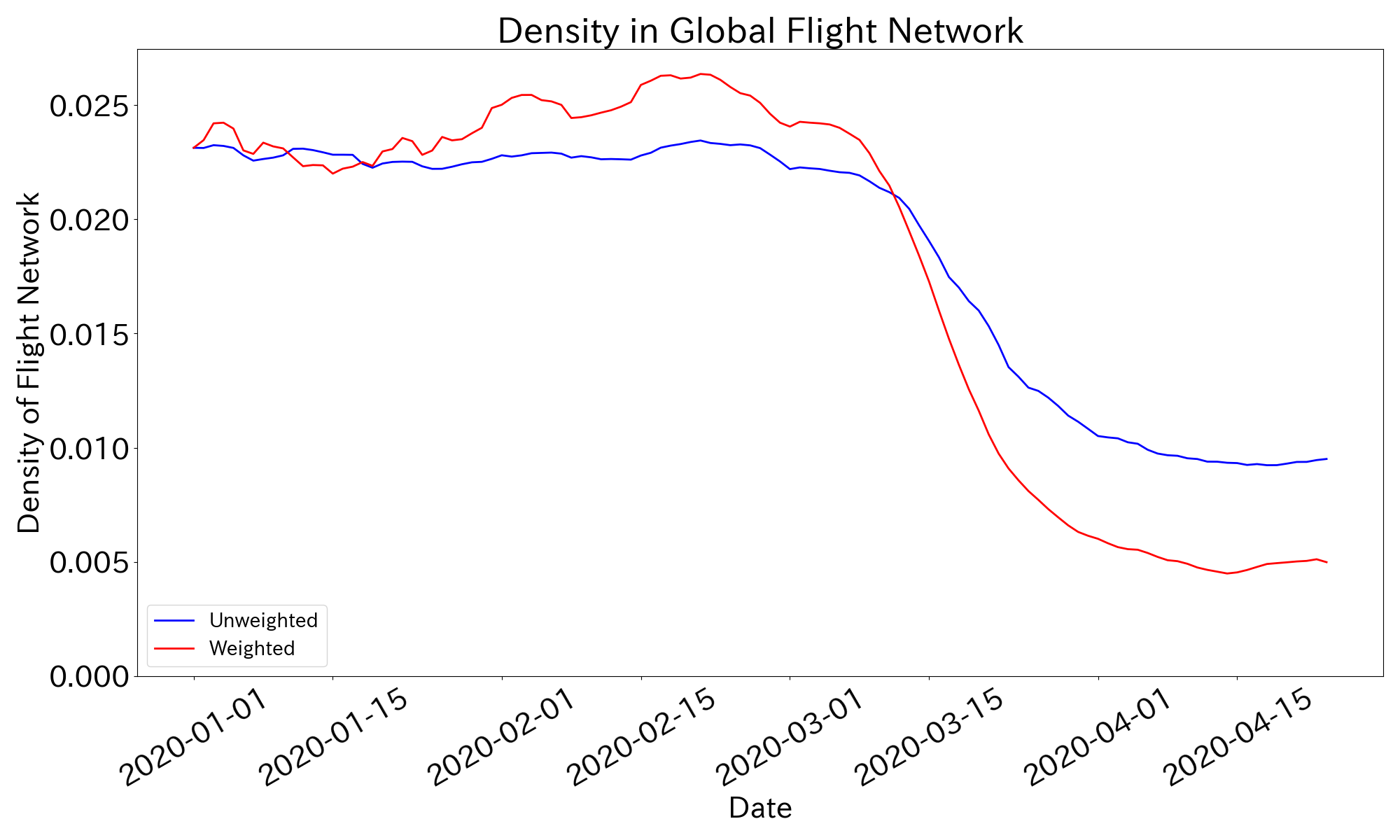}}
\caption{Density of Global Flight Network with Unweighted (Blue) and Weighted (Red) Flight Edges}
\label{fig:weight-global}
\end{figure}

\begin{figure}[htb]
\centerline{\includegraphics[width=\hsize]{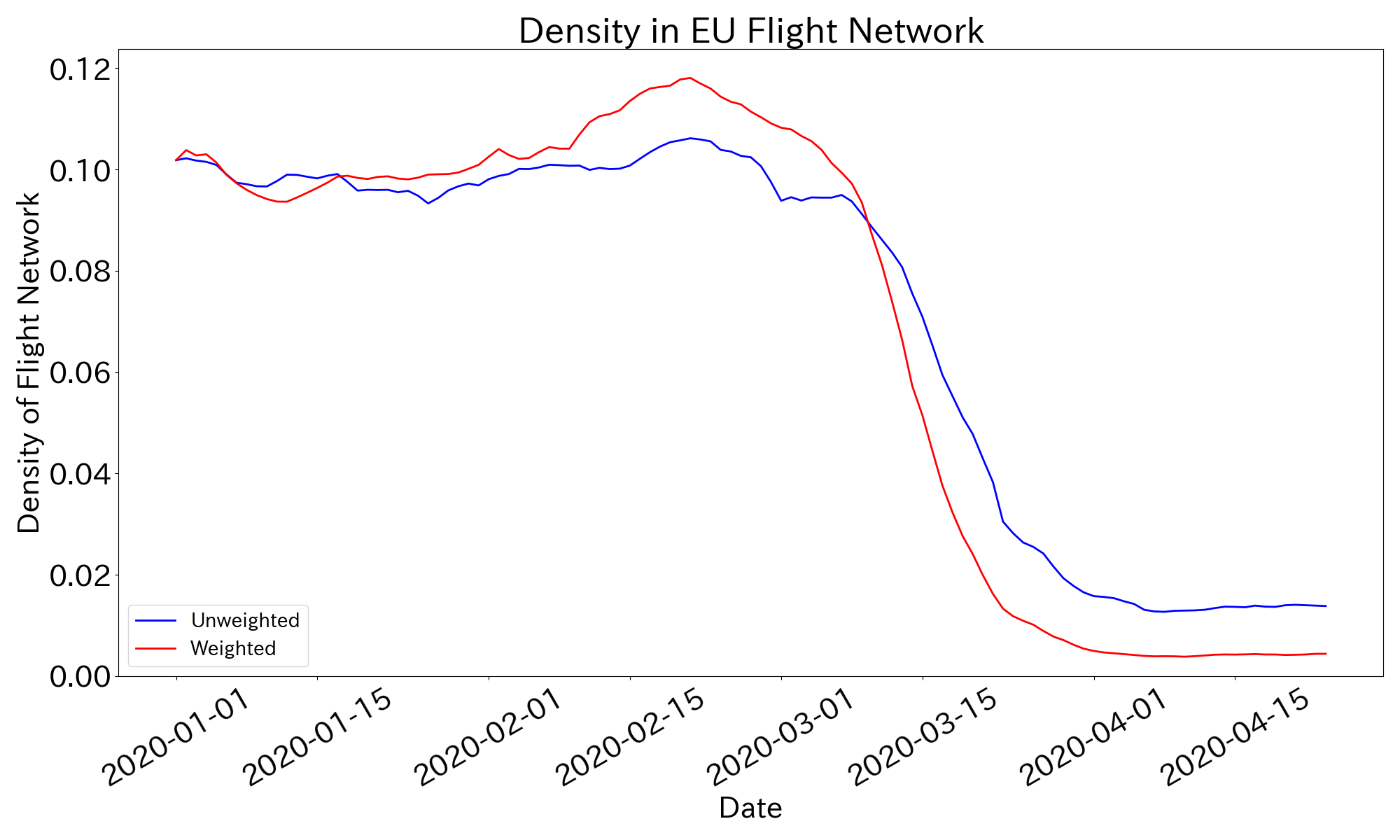}}
\caption{Density of Flight Network in Europe with Unweighted (Blue) and Weighted (Red) Flight Edges}
\label{fig:weight-eu}
\end{figure}

\begin{figure}[htb]
\centerline{\includegraphics[width=\hsize]{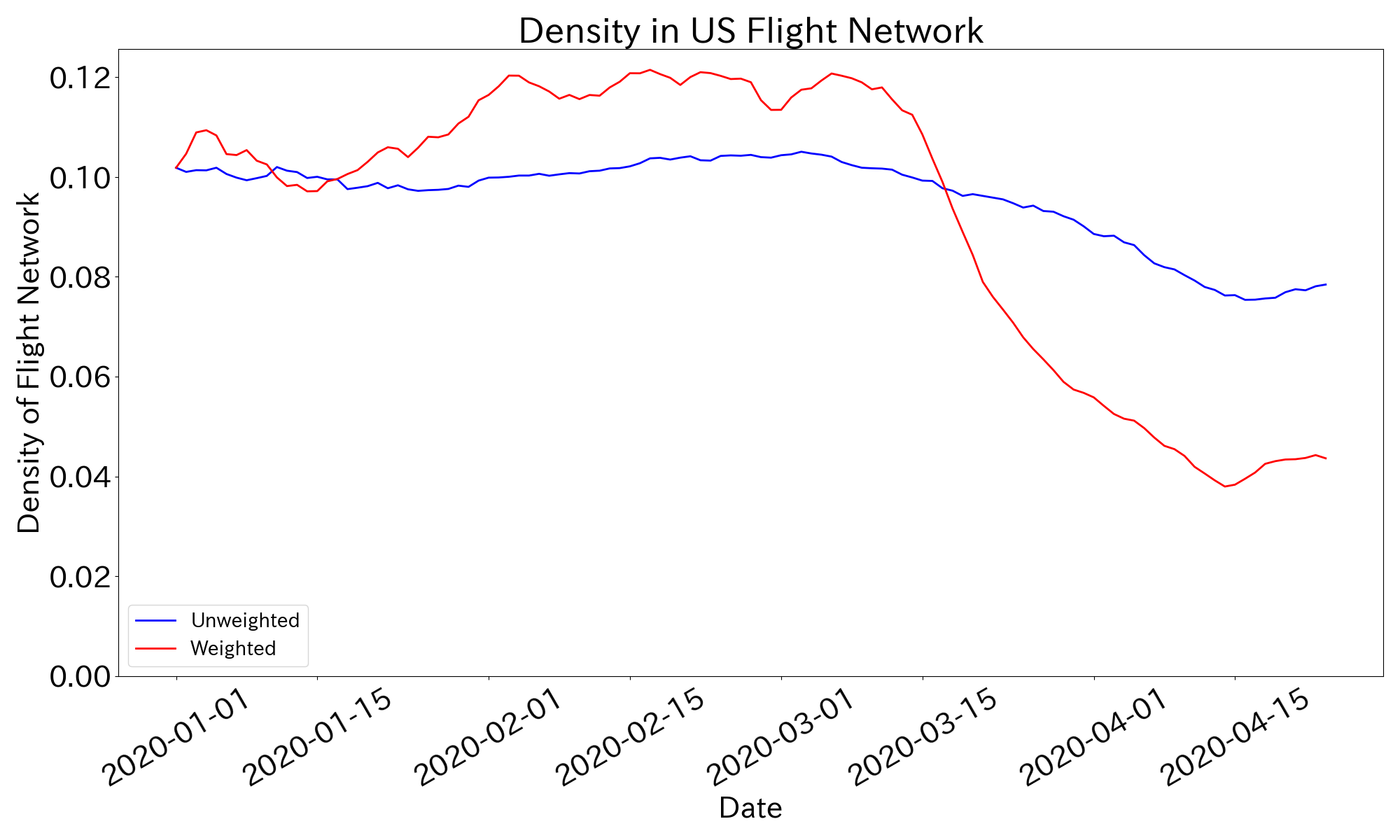}}
\caption{Density of Flight Network in the United States with Unweighted (Blue) and Weighted (Red) Flight Edges}
\label{fig:weight-us}
\end{figure}

The blue and red lines in Figure \ref{fig:weight-global} - \ref{fig:weight-us} represent the density of the daily flight network with the unweighted/weighted flight edges. The density of the whole flight network with weighted edges in Europe dropped to almost zero in April.

In the US, the density with weighted edges has dropped from 0.025 to 0.005 in March. Although the topology of the flight network remains unchanged, the frequency of domestic flights in the US significantly decreased as seen in other continents.

\subsection{Key Takeaways}

We applied some graph analytics to the flight data in order to evaluate the density of flight networks. Many graph metrics such as density, maximum SCC size, and diameter remarkably indicated that the flight networks in the overall world and Europe became more sparse in March. We also defined the "weighted" flight network density considering the frequency of flights, and showed the flight network in the United States also became more sparse from the mid of March to the mid of April.

\section{Impact on the Travel Industry in Barcelona}
\label{sec:barcelona}

Travel restrictions have a large impact on travel industries and tourism-dependent countries. In this section, we investigate the correlation and causal association between the decline of incoming flights, the number of infection cases, and the unemployment in Barcelona.

\subsection{Incoming Flight Reductions in Barcelona}

\begin{figure}[htb]
\centerline{\includegraphics[width=\hsize]{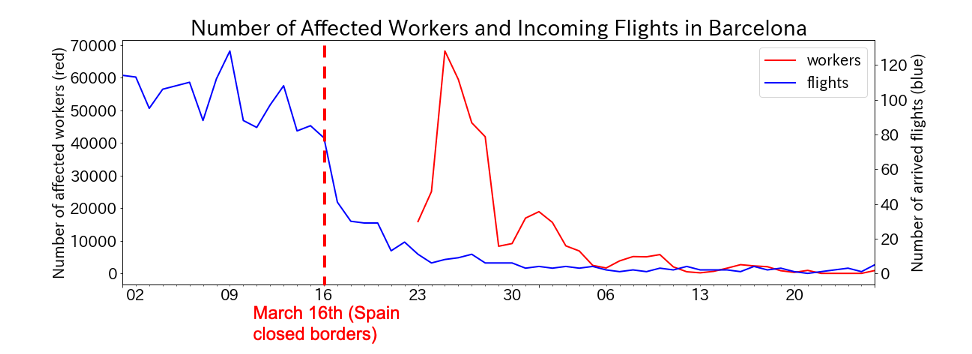}}
\caption{Changes of the Number of Affected Temporary Workers with the Number of Incoming Flights and Estimated Passengers in Barcelona}
\label{fig:bcn-workers}
\end{figure}

According to the unemployment dataset\cite{spain-workers}, the number of affected temporary workers (red line in Figure \ref{fig:bcn-workers}) gradually declined in April after it suddenly increased on Match 24. The number of flights (blue line) arriving in Barcelona gradually decreased until the beginning of April. These numbers seem to have some correlations, but more investigation is required to understand the causality and relationship between these numbers.

\subsection{Unemployment and New Infection Cases in Barcelona}

\begin{figure}[htb]
\centerline{\includegraphics[width=\hsize]{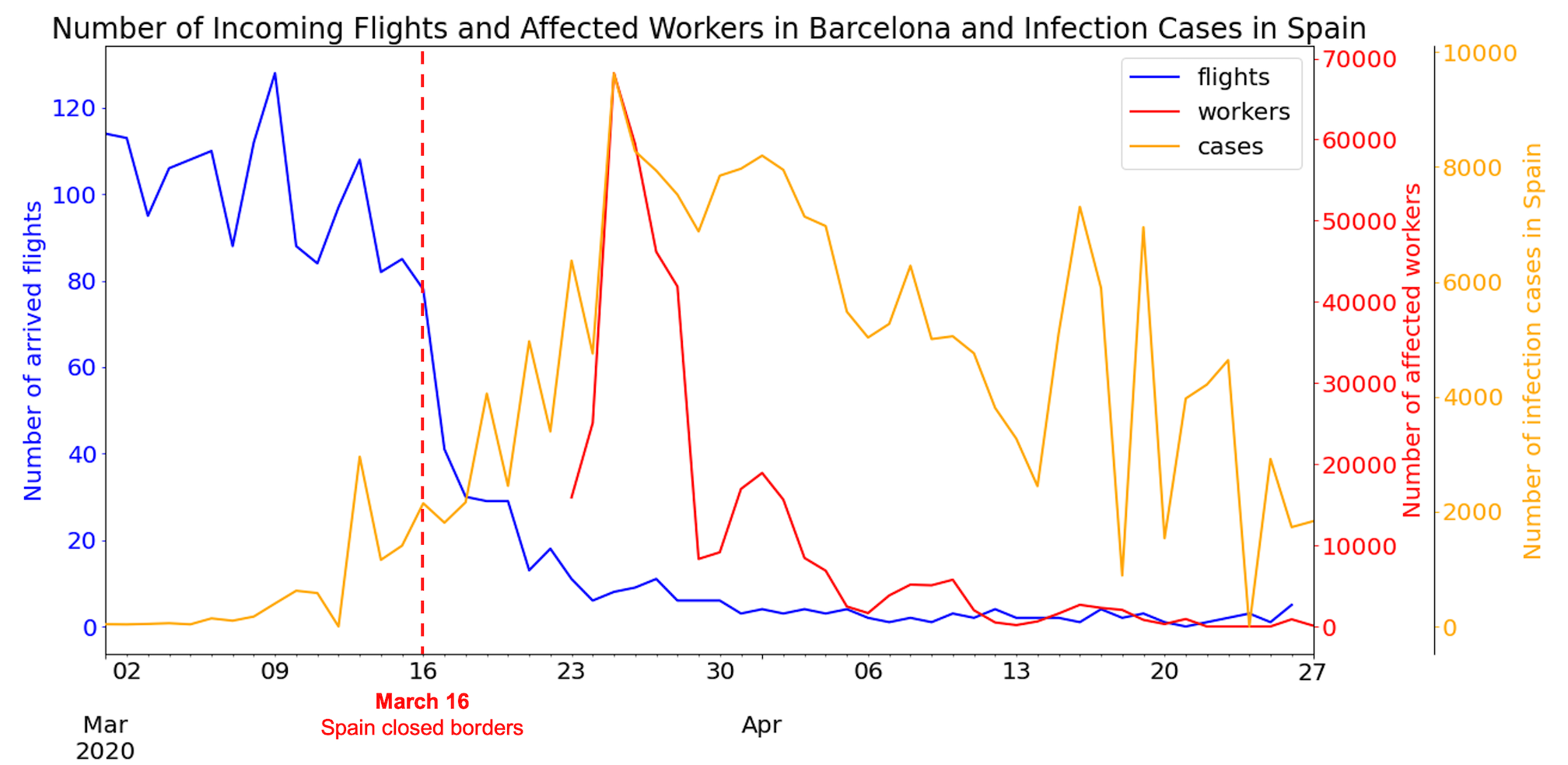}}
\caption{Changes of the Number of Affected Temporary Workers (red line) in Barcelona and the Number of New Infection Cases (orange line) in Spain}
\label{fig:bcn-cases}
\end{figure}

The number of flights to Barcelona (blue line in Figure \ref{fig:bcn-cases}) has been decreasing since the beginning of March, and suddenly halved on March 16 just before Spain closed its borders\cite{spain-close}.
Then, the number of affected temporary workers (red line) suddenly increased by 4.3 times (from 15,901 on March 23 to 68,170 on March 25). It is notable that the dataset regarding the number of affected workers became available from March 23. The number of new infection cases in Spain (orange line) gradually increased from March 12 and peaked at almost 10,000 on March 25.

\subsection{Flights to Barcelona and New Infection Cases in the Major Origin Countries}
\label{sec:new-cases}
Approximately half of all flights to Barcelona (BCN) come from France, Germany, and the UK. The government of Spain closed its borders on March 16, and the number of flights from these three countries decreased by half during  the following day. However, the number of new infection cases for these countries has been gradually increasing even before border closures. Considering the latent period of COVID-19 is commonly five to six days \cite{who-latent}, the reason the trend of increase of new infection cases continued for a week after the border closure may be due to the immigration of infected individuals before the border closure.

\begin{figure}[htb]
\centerline{\includegraphics[width=\hsize]{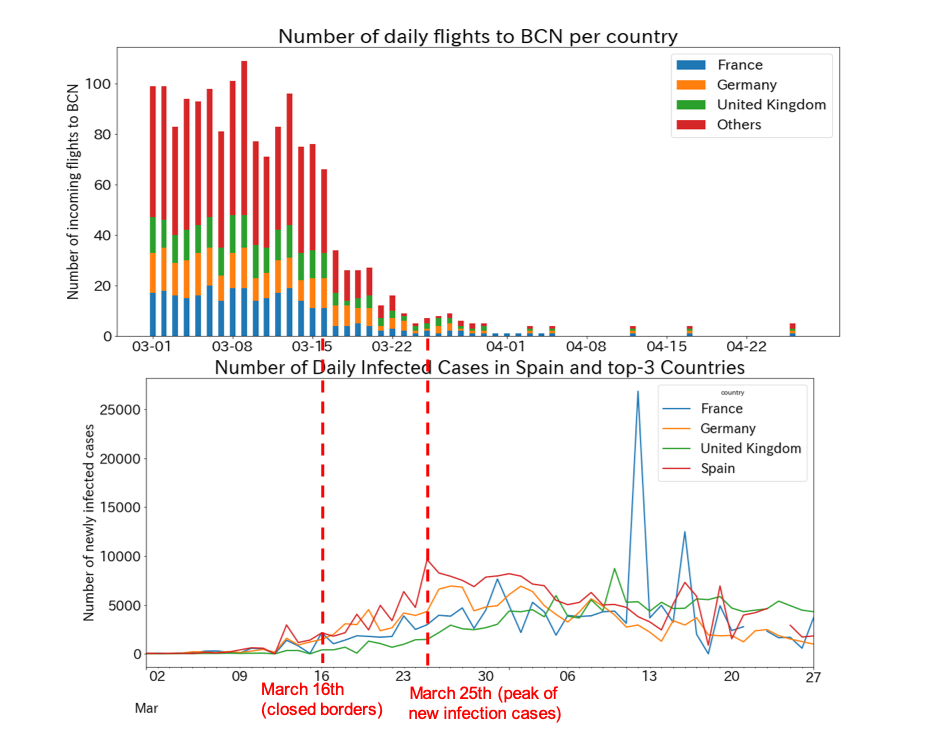}}
\caption{Number of Incoming Flights and New Infection Cases of the Top-3 Origin Countries}
\label{fig:bcn-flights}
\end{figure}

\subsection{Key Takeaways}
We compared the number of incoming flights, affected temporary workers, and the number of new infection cases in Barcelona to estimate how the travel restriction affected the tourism industry. We found the number of flights halved after Spain closed borders on March 16, and the number of affected workers jumped just after the government opened the application of unemployment on March 23. The number of new infection cases increased until March 25 after the border closures, potentially due to infections spreading during its latent period.

\section{Relationship Between Incoming Flights and Infection Cases}
\label{sec:cases}

\begin{figure}[htb]
\centerline{\includegraphics[width=\hsize]{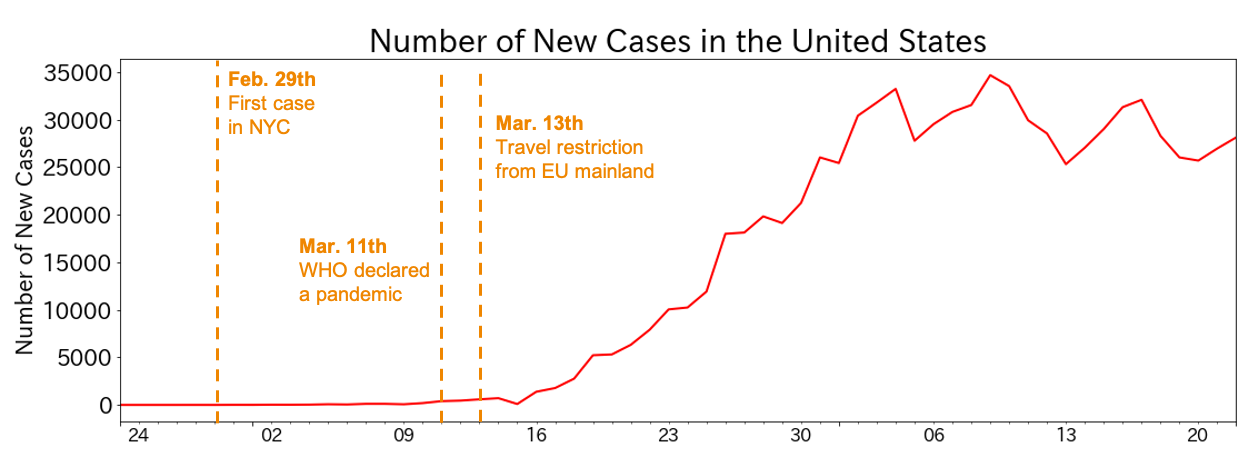}}
\caption{Number of New Infection Cases in the United States}
\label{fig:cases-us}
\end{figure}

A sequence analysis of COVID-19 virus found out that the virus type of the most infection cases in New York City is from Europe\cite{spread-nyc}. The first case was confirmed in the State of New York on February 29, and the number of infection cases gradually increased from March 15 (Figure \ref{fig:cases-us}). To investigate the effects of incoming flights from Europe, we compared the number of daily flights from Europe and new infection cases in the US.

\subsection{Number of Infection Cases in European Countries and the United States}
\begin{figure}[htb]
\centerline{\includegraphics[width=\hsize]{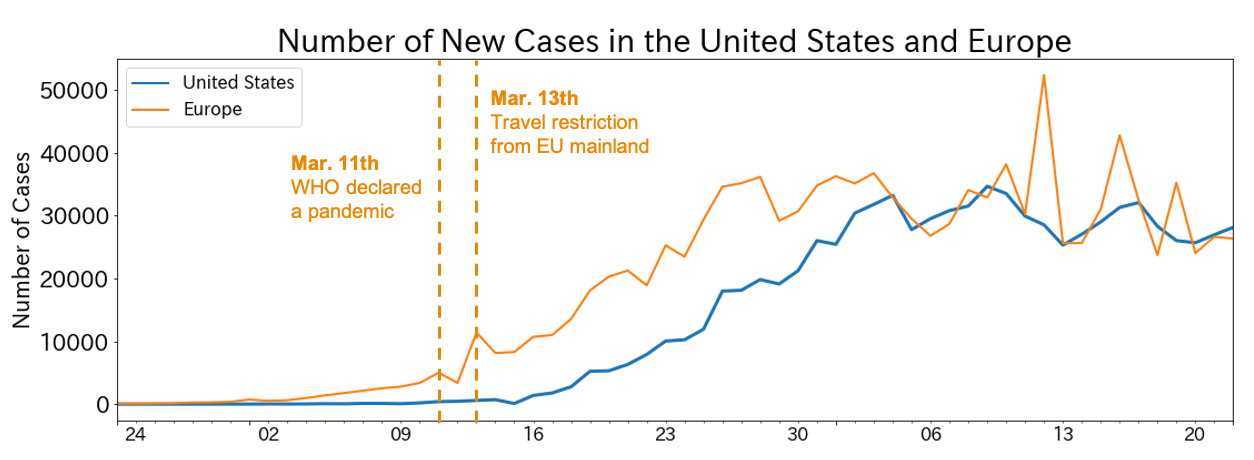}}
\caption{Comparison of the Total Number of Daily Infection Cases in the United States (bold blue line) and Europe (orange line)}
\label{fig:cases-us-eu-all}
\end{figure}

\begin{figure}[htb]
\centerline{\includegraphics[width=\hsize]{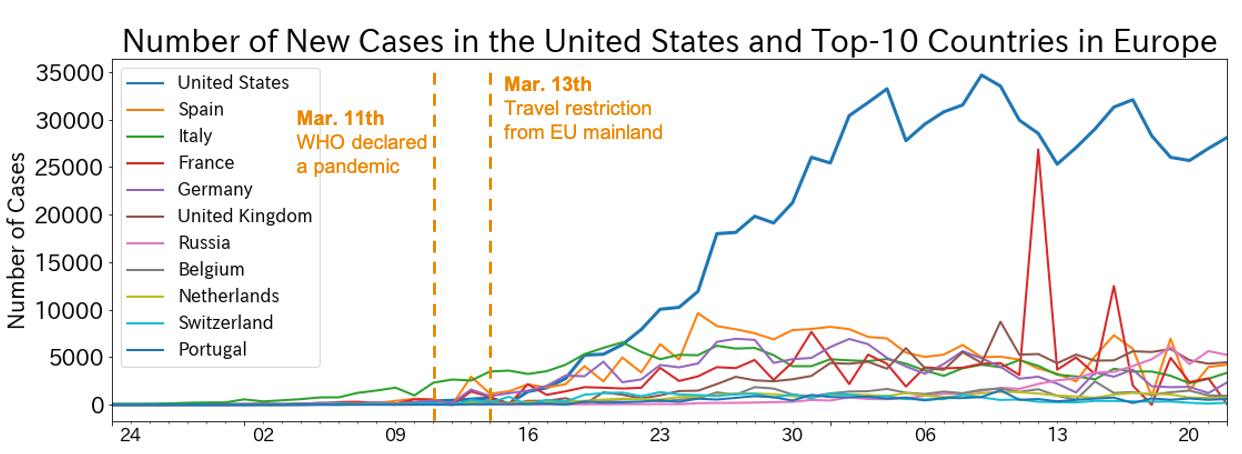}}
\caption{Comparison of the Number of Daily Infection Cases in the United States (bold blue line) and Top-10 Countries in Europe}
\label{fig:cases-us-eu-top10}
\end{figure}

The trend of increase in newly infected cases in Europe began a week before the US (Figure \ref{fig:cases-us-eu-all}). In particular, confirmed new infections in Italy started to increase in the beginning of March. Spain and Germany also confirmed cases before the US (Figure \ref{fig:cases-us-eu-top10}). The US announced to enforce travel restrictions from Europe on March 13 (EU mainland) and March 14 (the UK and Ireland). After these restrictions, infection cases gradually increased in the US from March 15 and exceeded 30,000 on April 2.

\subsection{Number of Incoming Flights from Europe and Infection Cases in United States}
\begin{figure}[htb]
\centerline{\includegraphics[width=\hsize]{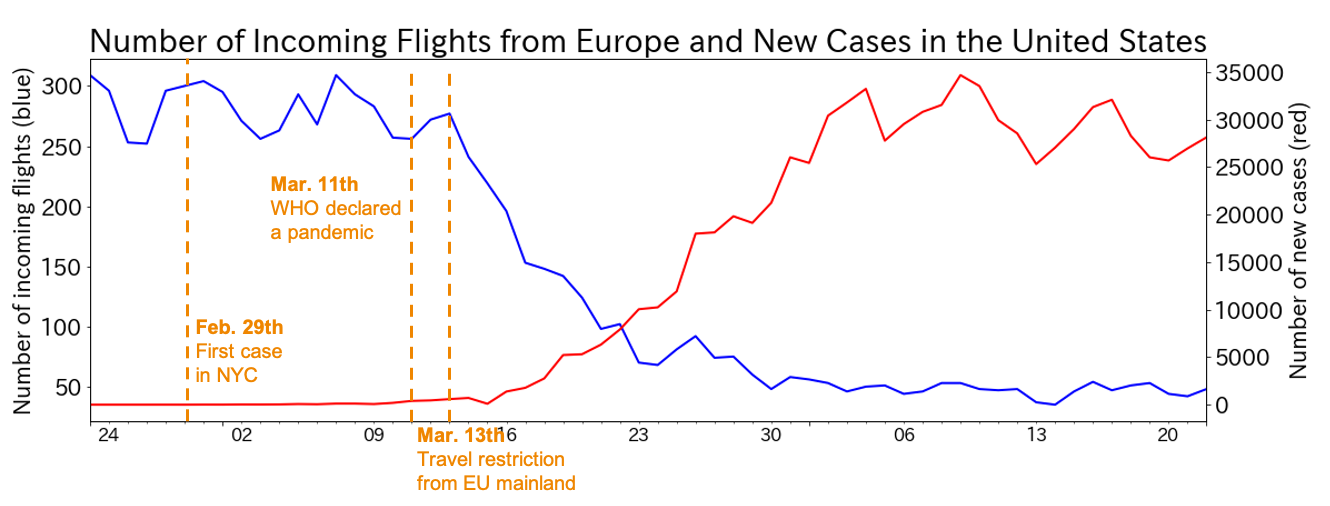}}
\caption{Number of Daily Incoming Flights from Europe and New Infection Cases in the United States}
\label{fig:flights-cases-us-eu}
\end{figure}

With the travel restrictions from European countries to the United States, the number of incoming flights (blue line in Figure \ref{fig:flights-cases-us-eu}) suddenly decreased from March 13 to March 24. However, the number of infection cases (red line in Figure \ref{fig:flights-cases-us-eu}) began to increase just after these restrictions. However, the latent period of COVID-19 is commonly five to six days as we mentioned in Section \ref{sec:new-cases}, therefore we cannot rule out the possibility that the infections had already spread out before the US enforced these restrictions.

\begin{figure}[htb]
\centerline{\includegraphics[width=\hsize]{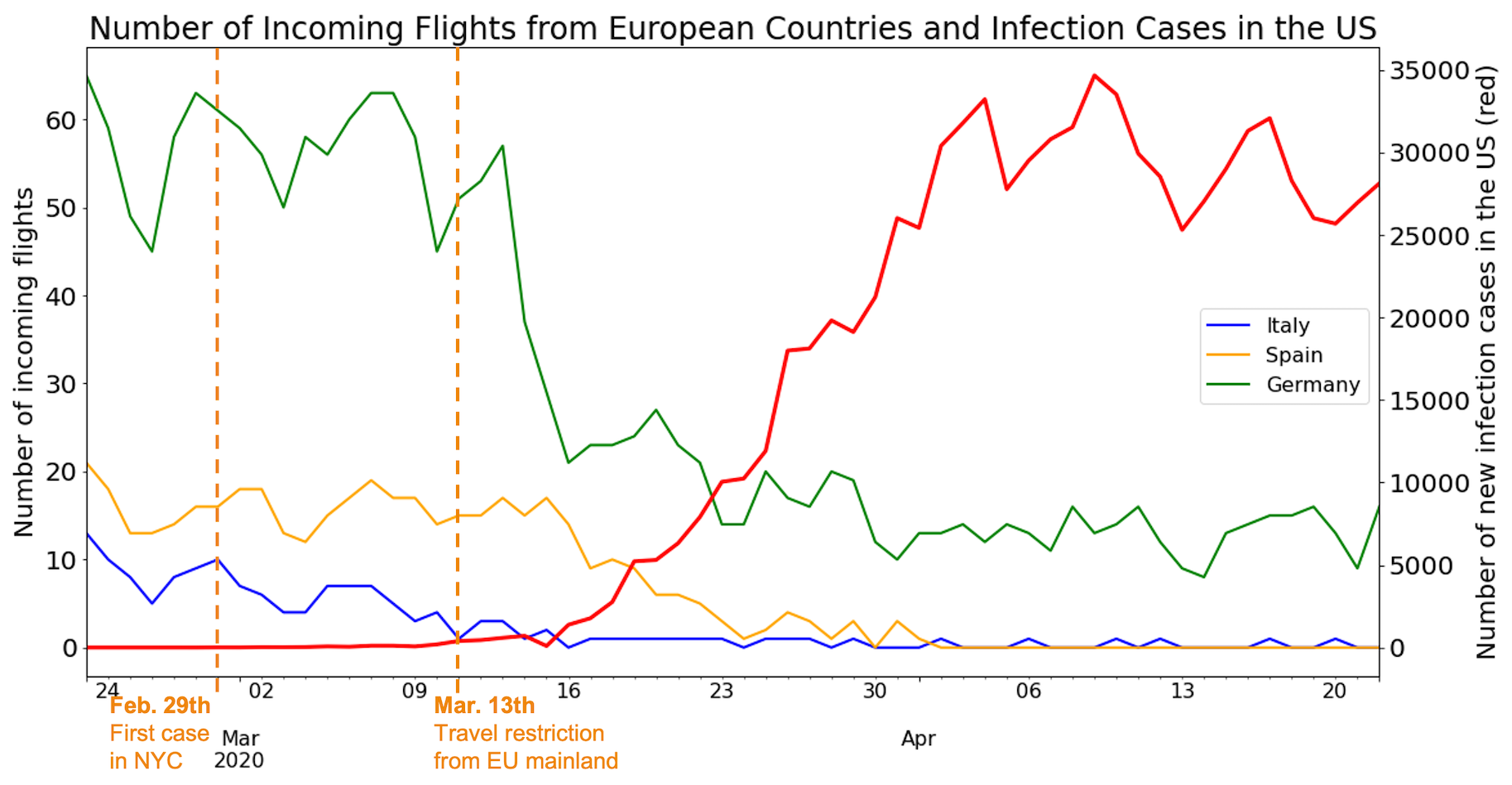}}
\caption{Number of Daily Incoming Flights from European Countries (Italy, Spain and Germany) and New Infection Cases in the United States}
\label{fig:flights-cases-eu}
\end{figure}

European countries such as Italy, Spain and Germany have already reported a certain number of infection cases before travel restrictions. Figure \ref{fig:flights-cases-eu} shows the number of daily international flights from these countries to the US and the number of daily new infections in the US (red lines). In Italy (blue line), more infection cases were reported than other European countries from March 1 (Figure \ref{fig:flights-cases-us-eu}), but the number of flights to the US became almost zero just after the travel restriction was enforced. On the other hand, some flights from Spain (orange line) and Germany (green line) to the US were still in operation, even after the restrictions were announced and infection cases in the US began to increase. These flights may be one of the reasons for the gradual increase of infections.

\section{Discussion}
\label{sec:discussion}

\subsection{Findings}
In Section 4, we found out when and how international flights in each continent and countries declined due to travel restrictions. The number of flights on April 1 had dropped to 10\% of that of regular seasons. The number of flights from Europe sharply decreased in the second half of March after the US announced the travel restriction from most of European countries on March 11 and the EU agreed to close borders on March 17. On the other hand, half of regular domestic flights in the US were still in operation at the end of March.

Besides counting the number of daily flights, we also applied several graph analytics to daily flight networks in order to explain how these networks became sparse in Section 5. With the result of analytics, we found out the density of the global flight network halved in March, and it has been divided into small strongly connected components from the mid of March to the end of April.

In Section 6, we evaluated the effect of reduction of incoming flights to Barcelona on the tourism industry (e.g. the number of affected temporary workers). The number of incoming flights dropped just after the travel restriction by the government of Spain on March 16, and the number of affected workers drastically increased after the government of Catalonia published the number of these workers from March 23. However, the number of infection cases still increased even after travel restrictions were announced.

We also investigated the relationship between incoming flights from Europe and the number of infections in the US in Section 7. We found that the trend of daily infections in Europe began a week ahead of the US, and that some European countries confirmed cases from March 1 while the number of infection cases in the US began to increase from March 15. We also found that  flights from Europe were still accepted in the US even after travel restrictions were announced on March 14.

\subsection{Related Work}
Many researches have conducted evaluations and estimations how COVID-19 infections affected international flights.

Stefano et al.\cite{IACUS2020104791} evaluated the economic impact of travel restrictions on flights. They constructed several scenarios of travel restrictions based on the past pandemic cases such as SARS in 2003 and MERS in 2005, and estimated the loss of global GDP may amount up to 1.98\% and that the number of unemployment could reach up to 5 millions.

Hien et al.\cite{LAU2020} evaluated the correlation of domestic COVID-19 cases and flight traffic volumes with the OpenSky Network dataset, and concluded the number of international flight routes and passengers have a correlation with the risk of COVID-19 exposures.

\subsection{Limitations and Possible Bias}
Since the flight dataset from the OpenSky Network is aggregated voluntarily, it has some limitations. First, some flight entries do not contain origin or destination airport codes. We excluded such entries before analysis to construct valid flight networks. Moreover, the dataset also has a significant regional bias. For example, many international and domestic flights connecting China, Oceania, South America, and some countries in Africa are missing.

\section{Conclusion}
\label{sec:conclusion}

\subsection{Overall Conclusion}
After the declaration of a pandemic by WHO on March 11 and government announcements of travel restrictions during the following week, the number of international flights around the world drastically declined, particularly in Europe. Although many domestic flights were still in operation in the US, the frequency gradually declined.

These travel restrictions affected tourism-dependent countries. In Barcelona in particular, many temporary workers made applications regarding their unemployment soon after the government opened applications on March 23. Moreover, the number of newly infected cases increased after travel restrictions were declared on March 16, most likely due to COVID-19’s longer latency period.

\subsection{Future Directions}
Due to the limitation of publicly available data, we could not conduct the same analysis for many countries as Barcelona in Section 6 for this paper. Therefore, we aim to analyze relationships among the number of flights, infection cases, and the impact on the economy for more countries and continents by combining various other data sets with the flight data in our next analysis. We also aim to focus on counties where the number of infections has soared, and investigate how the increase of cases and reduction of flights affect the economy.

We also plan to trace and predict viral mutations with the flight network data. The type of COVID-19 virus varies from region to region\cite{TreeTime, Nextstrain}, and this may provide insight to analyzing the route of infections with viral types. With the daily flight network data, we hope to estimate the date and route of infections more precisely.

In order to predict the economic impacts such as propagations of revenue declines in travel industries for each region, we will apply advanced machine learning techniques for time-series graph data such as Graph Convolutional Network (GCN). We will also evaluate the importance of flight network features as well as demographic features of regions (countries and continents) in our next analysis.


\bibliographystyle{IEEEtran}
\bibliography{IEEEabrv,references}

\begin{thebibliography}{10}
\providecommand{\url}[1]{#1}
\csname url@samestyle\endcsname
\providecommand{\newblock}{\relax}
\providecommand{\bibinfo}[2]{#2}
\providecommand{\BIBentrySTDinterwordspacing}{\spaceskip=0pt\relax}
\providecommand{\BIBentryALTinterwordstretchfactor}{4}
\providecommand{\BIBentryALTinterwordspacing}{\spaceskip=\fontdimen2\font plus
\BIBentryALTinterwordstretchfactor\fontdimen3\font minus
  \fontdimen4\font\relax}
\providecommand{\BIBforeignlanguage}[2]{{%
\expandafter\ifx\csname l@#1\endcsname\relax
\typeout{** WARNING: IEEEtran.bst: No hyphenation pattern has been}%
\typeout{** loaded for the language `#1'. Using the pattern for}%
\typeout{** the default language instead.}%
\else
\language=\csname l@#1\endcsname
\fi
#2}}
\providecommand{\BIBdecl}{\relax}
\BIBdecl

\bibitem{whitehouse-airlines}
T.~W. House, ``Remarks by president trump and vice president pence at a
  coronavirus briefing with airline ceos,''
  \url{https://www.whitehouse.gov/briefings-statements/remarks-president-trump-vice-president-pence-coronavirus-briefing-airline-ceos/},
  2020.

\bibitem{tourism-gdp}
TCdata360, ``Travel and tourism direct contribution to gdp,''
  \url{https://www.whitehouse.gov/briefings-statements/remarks-president-trump-vice-president-pence-coronavirus-briefing-airline-ceos/},
  2020.

\bibitem{barcelona-tourism}
B.~C. Council, ``Barcelona in numbers: Tourism, between wealth and residents’
  complaints | barcelona metròpolis,''
  \url{https://www.barcelona.cat/metropolis/en/contents/tourism-between-wealth-and-residents-complaints},
  2020.

\bibitem{opensky-data}
\BIBentryALTinterwordspacing
M.~{Schäfer}, M.~{Strohmeier}, V.~{Lenders}, I.~{Martinovic}, and
  M.~{Wilhelm}, ``Bringing up opensky: A large-scale ads-b sensor network for
  research,'' in \emph{IPSN-14 Proceedings of the 13th International Symposium
  on Information Processing in Sensor Networks}, 2014, pp. 83--94. [Online].
  Available: \url{https://opensky-network.org}
\BIBentrySTDinterwordspacing

\bibitem{opensky-web}
T.~O. Network, ``The opensky network - free ads-b and mode s data for
  research,'' \url{https://opensky-network.org/}, 2020.

\bibitem{aircraft-passengers}
``List of aircraft type designators - wikipedia.''
  \url{https://en.wikipedia.org/wiki/List\_of\_aircraft\_type\_designators},
  2020.

\bibitem{national-emergency}
``Proclamation on declaring a national emergency concerning the novel
  coronavirus disease (covid-19) outbreak,''
  \url{https://www.whitehouse.gov/presidential-actions/proclamation-declaring-national-emergency-concerning-novel-coronavirus-disease-covid-19-outbreak/},
  2020.

\bibitem{us-close-eu}
``Proclamation - suspension of entry as immigrants and nonimmigrants of certain
  additional persons who pose a risk of transmitting 2019 novel coronavirus,''
  \url{https://www.whitehouse.gov/presidential-actions/proclamation-suspension-entry-immigrants-nonimmigrants-certain-additional-persons-pose-risk-transmitting-2019-novel-coronavirus/},
  2020.

\bibitem{us-close-canada}
``President donald j. trump is taking necessary safety measures at the border
  to prevent further spread of the coronavirus,''
  \url{https://www.whitehouse.gov/briefings-statements/president-donald-j-trump-taking-necessary-safety-measures-border-prevent-spread-coronavirus/},
  2020.

\bibitem{germany-border}
``Coronavirus: Germany to impose border controls over coronavirus,''
  \url{https://www.bbc.com/news/world-europe-51897069}, 2020.

\bibitem{eu-close}
E.~Council", ``Video conference of the members of the european council, 17
  march 2020,''
  \url{https://www.consilium.europa.eu/en/meetings/european-council/2020/03/17/},
  2020.

\bibitem{wb-passengers}
\BIBentryALTinterwordspacing
T.~W. Bank", ``Air transport, passengers carried,'' 2020. [Online]. Available:
  \url{https://data.worldbank.org/indicator/IS.AIR.PSGR}
\BIBentrySTDinterwordspacing

\bibitem{us-flights}
\BIBentryALTinterwordspacing
L.~Josephs", ``Coronavirus forces airlines to consider a once unthinkable
  possibility - halting us flights,'' 2020. [Online]. Available:
  \url{https://www.cnbc.com/2020/03/16/coronavirus-makes-airlines-consider-chances-for-a-halt-to-us-flights.html}
\BIBentrySTDinterwordspacing

\bibitem{spain-workers}
\BIBentryALTinterwordspacing
G.~de~Catalunya, ``Temporary employment regulation files (erte) presented daily
  and number of affected workers.'' 2020. [Online]. Available:
  \url{https://analisi.transparenciacatalunya.cat/en/Treball/Evoluci-di-ria-dels-Expedients-de-Regulaci-Tempora/atmi-6snp}
\BIBentrySTDinterwordspacing

\bibitem{spain-close}
\BIBentryALTinterwordspacing
L.~Moncloa, ``Government restricts access to travellers at spain's external
  borders,'' 2020. [Online]. Available:
  \url{https://www.lamoncloa.gob.es/lang/en/gobierno/news/Paginas/2020/20200322travellers.aspx}
\BIBentrySTDinterwordspacing

\bibitem{who-latent}
\BIBentryALTinterwordspacing
W.~H. Organization", ``"coronavirus disease 2019 - world health organization."
  q\&a on coronaviruses (covid-19),'' 2020. [Online]. Available:
  \url{https://www.who.int/emergencies/diseases/novel-coronavirus-2019/question-and-answers-hub/q-a-detail/q-a-coronaviruses}
\BIBentrySTDinterwordspacing

\bibitem{spread-nyc}
A.~S. Gonzalez-Reiche, M.~M. Hernandez, M.~J. Sullivan, B.~Ciferri,
  H.~Alshammary, A.~Obla, S.~Fabre, G.~Kleiner, J.~Polanco, Z.~Khan
  \emph{et~al.}, ``Introductions and early spread of sars-cov-2 in the new york
  city area,'' \emph{Science}, 2020.

\bibitem{IACUS2020104791}
S.~M. Iacus, F.~Natale, C.~Santamaria, S.~Spyratos, and M.~Vespe, ``Estimating
  and projecting air passenger traffic during the covid-19 coronavirus outbreak
  and its socio-economic impact,'' \emph{Safety Science}, vol. 129, p. 104791,
  2020.

\bibitem{LAU2020}
H.~Lau, V.~Khosrawipour, P.~Kocbach, A.~Mikolajczyk, H.~Ichii, M.~Zacharski,
  J.~Bania, and T.~Khosrawipour, ``The association between international and
  domestic air traffic and the coronavirus (covid-19) outbreak,'' \emph{Journal
  of Microbiology, Immunology and Infection}, 2020.

\bibitem{TreeTime}
\BIBentryALTinterwordspacing
P.~Sagulenko, V.~Puller, and R.~A. Neher, ``{TreeTime: Maximum-likelihood
  phylodynamic analysis},'' \emph{Virus Evolution}, vol.~4, no.~1, 01 2018,
  vex042. [Online]. Available: \url{https://doi.org/10.1093/ve/vex042}
\BIBentrySTDinterwordspacing

\bibitem{Nextstrain}
\BIBentryALTinterwordspacing
J.~Hadfield, C.~Megill, S.~M. Bell, J.~Huddleston, B.~Potter, C.~Callender,
  P.~Sagulenko, T.~Bedford, and R.~A. Neher, ``{Nextstrain: real-time tracking
  of pathogen evolution},'' \emph{Bioinformatics}, vol.~34, no.~23, pp.
  4121--4123, 05 2018. [Online]. Available:
  \url{https://doi.org/10.1093/bioinformatics/bty407}
\BIBentrySTDinterwordspacing

\end{thebibliography}

\end{document}